\documentclass[twocolumn,useAMS,usenatbib]{mnras} \usepackage{natbib}

\usepackage{amsmath,latexsym,amssymb} \usepackage{epsfig,graphicx} \topmargin-1cm
\usepackage{epstopdf}
\usepackage{float}
\usepackage{fixltx2e}
\usepackage{color}
\usepackage{url}
\graphicspath{{figures/}}
\DeclareGraphicsExtensions{.pdf,.png,.jpg,.eps}

\def\reff@jnl#1{{\rm#1\/}}

\def\aj{\reff@jnl{AJ}}                  
\def\araa{\reff@jnl{ARA\&A}}            
\def\apj{\reff@jnl{ApJ}}                        
\def\apjl{\reff@jnl{ApJ}}               
\def\apjs{\reff@jnl{ApJS}}              
\def\apss{\reff@jnl{Ap\&SS}}            
\def\aap{\reff@jnl{A\&A}}               
\def\aapr{\reff@jnl{A\&A~Rev.}}         
\def\aaps{\reff@jnl{A\&AS}}             
\def\baas{\reff@jnl{BAAS}}              
\def\jcap{\reff@jnl{JCAP}}              
\def\jrasc{\reff@jnl{JRASC}}            
\def\memras{\reff@jnl{MmRAS}}           
\def\mnras{\reff@jnl{MNRAS}}            
\def\physrep{\reff@jnl{Phys.Rep.}}
\def\pra{\reff@jnl{Phys.Rev.A}}         
\def\prb{\reff@jnl{Phys.Rev.B}}         
\def\prc{\reff@jnl{Phys.Rev.C}}         
\def\prd{\reff@jnl{Phys.Rev.D}}         
\def\prl{\reff@jnl{Phys.Rev.Lett}}      
\def\pasp{\reff@jnl{PASP}}              
\def\pasj{\reff@jnl{PASJ}}              
\def\skytel{\reff@jnl{S\&T}}            
\def\solphys{\reff@jnl{Solar~Phys.}}    
\def\sovast{\reff@jnl{Soviet~Ast.}}     
\def\ssr{\reff@jnl{Space~Sci.Rev.}}     
\def\nat{\reff@jnl{Nature}}             

\newcommand{\hmpc}{\ensuremath{h^{-1}\mathrm{Mpc}}}
\newcommand{\hkpc}{\ensuremath{h^{-1}\mathrm{kpc}}}
\newcommand{\hMsun}{\ensuremath{h^{-1}M_{\odot}}}

\newcommand{\beq}{\begin{equation}}
\newcommand{\eeq}{\end{equation}}
\newcommand{\beqa}{\begin{eqnarray}}
\newcommand{\eeqa}{\end{eqnarray}}

\newcommand*{\email}[1]{\href{mailto:#1}{#1}}

\title[Group-scale intrinsic galaxy alignments]{ Group-scale intrinsic galaxy alignments in the Illustris-TNG and MassiveBlack-II simulations} 
\author[A.~Tenneti et al.]{
Ananth Tenneti$^{1}$\thanks{E-mail: \email{a.tenneti@ucl.ac.uk}}, 
Thomas D. Kitching$^{1}$, 
Benjamin Joachimi$^{2}$, 
Tiziana Di Matteo$^{3}$
\\
$^{1}$Mullard Space Science Laboratory, University College London, Holmbury St Mary, Dorking, Surrey RH5 6NT, UK\\
$^{2}$Department of Physics and Astronomy, University College London, Gower Street, London WC1E 6BT, UK\\
$^{3}$McWilliams Center for Cosmology, Dept. of Physics, Carnegie Mellon University,
Pittsburgh PA 15213, USA\\
}

\begin{document}
\maketitle

\begin{abstract}
  We study the alignments of satellite galaxies, and their anisotropic distribution, with respect to location and orientation of their host central galaxy in MassiveBlack-II and IllustrisTNG simulations. We find that: the shape of the satellite system in halos of mass ($> 10^{13}h^{-1}M_{\odot}$) is well aligned with the shape of the central galaxy at $z=0.06$ with the mean alignment between the major axes being 
  $\sim \Delta \theta = 12^{\circ}$ when compared to a uniform random distribution; that satellite galaxies tend to be anisotropically distributed along the major axis of the central galaxy with a stronger alignment in halos of higher mass or luminosity; and that the satellite distribution is more anisotropic for central galaxies with lower star formation rate, which are spheroidal, and for red central galaxies.
  Radially we find that satellites tend to be distributed along the major axis of the shape of the stellar component of central galaxies at smaller scales and the dark matter component on larger scales. We find that the dependence of satellite anisotropy on central galaxy properties and the radial distance is similar in both the simulations with a larger amplitude in MassiveBlack-II. The orientation of satellite galaxies tends to point toward the location of the central galaxy at small scales and this correlation decreases with increasing distance, and the amplitude of satellite alignment is higher in high mass halos. However, the projected ellipticities do not exhibit a scale-dependent radial alignment, as has been seen in some observational measurements.
\end{abstract}


\section{Introduction} \label{S:intro}
The intrinsic alignment (IA) of galaxies is a significant systematic in the science analysis of upcoming weak lensing surveys \citep{{2012MNRAS.424.1647K},{2016MNRAS.456..207K}}. IAs are also of interest as a probe of the physics of galaxy formation and evolution. The next generation ground and space-based telescope surveys such as the Large Synoptic Survey Telescope \footnote{\url{https://www.lsst.org/lsst/}} (LSST;\citealt{2009arXiv0912.0201L}), \emph{Euclid} \footnote{\url{http://sci.esa.int/euclid/}, \url{https://www.euclid-ec.org/}} \citealt{2011arXiv1110.3193L} and Wide-Field Infrared Survey Telescope \footnote{\url{https://wfirst.gsfc.nasa.gov/}} (WFIRST; \citealt{2015arXiv150303757S}) aim to employ weak lensing to constrain cosmological parameters to sub-percent level precision. However, it is well known that IAs are a contaminant in weak lensing measurements, and not taking the effect of IA into account can lead to a bias in the constraints of cosmological parameters such as the dark energy equation of state of up to $\sim 70 \%$ \citep{2016MNRAS.456..207K}. However, the effect of IAs can be mitigated by marginalizing over nuisance parameters in analytical models of IA \citep{{2010A&A...523A...1J},{2012JCAP...05..041B}}. The most straightforward way in which IAs have been modelled analytically is through the linear alignment model \citep{{Catelan2001},{PhysRevD.70.063526}}, and extensions of it to non-linear scales \citep{{Bridle_2007,Blazek_2015}}. The radial scaling predicted by the models has been found to be approximately consistent with observational measurements on linear scales, and the modelling can be extended up to the mildly non-linear regime \citep{{2011A&A...527A..26J},{2015MNRAS.450.2195S}}. However, on smaller scales this modelling becomes less physically motivated and so instead halo model prescriptions have been proposed to model IAs. A small scale halo model of IA was developed by \cite{Schneider2010}. This model assumes that the satellite galaxies are distributed symmetrically around the central galaxy and are oriented towards the central. In this paper, we test these assumptions on small scale galaxy alignments in the MassiveBlack-II \citep{2015MNRAS.450.1349K} and IllustrisTNG \citep{{2019ComAC...6....2N},{2018MNRAS.475..648P},{2018MNRAS.475..676S},{2018MNRAS.475..624N},{2018MNRAS.477.1206N},{2018MNRAS.480.5113M}} cosmological hydrodynamic simulations. In particular, we study the distribution and orientation of satellite galaxies with respect to the location and orientation of their host central galaxy. 

Cosmological simulations of galaxy formation have emerged as a useful tool to study IA and baryonic effects on weak lensing. For example IAs have been studied in simulations such as MassiveBlack-II \citep{2015MNRAS.450.1349K}, Horizon-AGN \citep{2014MNRAS.444.1453D}, EAGLE \citep{2015MNRAS.446..521S} and Illustris \citep{{2014Natur.509..177V},{2014MNRAS.444.1518V},{2014MNRAS.445..175G}}. The measurements of IA correlation functions from simulations have been found to be in good agreement with observations and broadly in agreement with each other, in particular results as a function of properties such as galaxy mass, luminosity and also the radial scaling of the alignment signal \citep{{2015MNRAS.448.3522T},{2015MNRAS.454.3328V},{2015MNRAS.454.2736C},{2017MNRAS.468..790H}}. It has to be noted though that some differences pertaining to tangential alignment of disk galaxies have been identified across simulations \citep{{2015MNRAS.454.2736C},{2016MNRAS.462.2668T}}. \cite{2019arXiv190109925S} used the MassiveBlack-II and Illustris simulations to consider the impact of the assumption of the spherically symmetric distribution of satellite galaxies used in the halo model of \cite{Schneider2010} for small scale intrinsic alignments modelling. \cite{2019arXiv190109925S} found that the anisotropic satellite distribution seen in the simulations leads to biases on the cosmological parameter constraints from weak lensing when alignments are modelled using the halo model. \cite{2016MNRAS.460.3772S} used the EAGLE simulation to study the alignment of the central galaxy with the shape traced by the satellites in the halo. \cite{2015arXiv151200400W} studied the distribution of satellites in the plane of central galaxy and with respect to filaments in the Horizon-AGN simulation. \cite{2014ApJ...791L..33D} used hydrodynamical simulations which includes gas cooling, star formation and feedback to investigate satellite anisotropy and found that it is stronger in red central galaxies. 
The anisotropic distribution of satellite galaxies with respect to shape of central galaxies has been studied observationally in various studies \citep[e.g.,][]{{2005ApJ...628L.101B},{2006MNRAS.369.1293Y},{2010ApJ...709.1321A},{2012ApJ...752...99N}}. \cite{2005ApJ...628L.101B} found that satellites are anisotropically distributed along the major axis of the central gaalxy in SDSS and that the degree of anisotropy decreases with distance from the central. \cite{2006MNRAS.369.1293Y} similary used galaxy groups from SDSS and found satellite anisotropy with more stronger anisotropy in red central galaxies and massive halos. \cite{2010ApJ...709.1321A} found that red, high mass central galaxies with low star formation rates have stronger satellite anisotropy using isolated galaxies in SDSS. \cite{2012ApJ...752...99N} found that the satellites in early type host galaxies are more anisotropically distributed along the central galaxy's major axis when compared with late-type galaxies using the data from COSMOS survey. Galaxy alignments in groups have also been studied through observations. \cite{{2016MNRAS.463..222H},{2018MNRAS.474.4772H}} used the redMaPPer cluster catalog data to study the central galaxy alignment with the satellite distribution and the radial alignment of satellite galaxies with respect to groups. More recently, \cite{2019arXiv190500370G} used the GAMA+KiDS data to study the scale dependence of satellite alignments with respect to the central galaxy. In this paper, we measure the alignments of the shape of central galaxy with the shape of the satellite system to compare against earlier simulation results. Measurements in MassiveBlack-II and IllustrisTNG simulations will also help to explore any differences in satellite galaxy alignments due to baryonic feedback models. We also present the scale dependent radial alignments of satellite shapes for comparison with the observational measurements of galaxy alignments in groups \citep{2019arXiv190500370G}.

The rest of the paper is organized as follows. In Section~\ref{S:methods}, we provide the details of MassiveBlack-II and Illustris simulation along with the methods adopted in the paper. In Section~\ref{S:results}, we present the results on the alignments of the central galaxy shape with the shape of the satellite system and the distribution of satellite galaxies along the central galaxy. We explore the dependence on various galaxy properties such as mass, luminosity, color and morphological type. In this section, we also present the small-scale galaxy alignment correlation functions in $3\rm{D}$, as well as projected correlation functions for comparison with observations. Finally, we present conclusions in Section~\ref{S:conclusions}. 

\section{Methods} \label{S:methods}
In this paper, we analyze the anisotropic distribution of satellites in the Illustris-TNG and MassiveBlack-II simulations. The details of the simulations are described below. 

\subsection{Simulations: Illustris-TNG and MassiveBlack-II} \label{S:sims}

The MassiveBlack-II (MB-II) simulation \citep{2015MNRAS.450.1349K} is a high resolution, cosmological hydrodynamic simulation performed with the {\sc p-gadget} code (a hybrid version of the smoothed-particle hydrodynamics (SPH) code, {\sc gadget2} \citep{2005MNRAS.361..776S} which has been upgraded to run on Petaflop scale supercomputers). The simulation is evolved in a cubic periodic box of volume $(100\hmpc)^3$ with $N_\mathrm{part} = 2\times 1792^{3}$ dark matter and gas
particles. The resolution is such that the mass of a single dark matter particle is $m_\text{DM} =
1.1\times 10^{7}\hMsun$ and the initial mass of a gas particle is
$m_\text{gas} = 2.2\times 10^{6}\hMsun$. The mass of each star
particle is $m_\text{star} = 1.1\times 10^{6}\hMsun$ and the gravitational smoothing length is $\epsilon = 1.85$\hkpc\ in
comoving units. The cosmological
parameters used in the simulation are based on \emph{WMAP7} \citep{2011ApJS..192...18K}. In addition to gravity and hydrodynamics, the simulation also includes the physics of multiphase ISM model
with star formation \citep{2003MNRAS.339..289S}, black hole accretion
and feedback
\citep{2005MNRAS.361..776S,2012ApJ...745L..29D}. We refer to \cite{2015MNRAS.450.1349K} for a more detailed description of the MB-II simulation and baryonic feedback models.

The IllustrisTNG simulation is performed with the {\sc AREPO TreePM} moving-mesh code
\citep{2010MNRAS.401..791S}  in a box of volume $(75\hmpc)^3$. The galaxy formation physics includes subgrid-model for star formation and associated supernova feedback, black hole accretion and feedback, stellar wind feedback, primordial and metal-line radiative cooling and magnetic fields. A detailed description of the models can be found in \cite{2017MNRAS.465.3291W} and \cite{2018MNRAS.473.4077P}. The simulation follows $2 \times 1820^3$ dark
matter particles and gas particles with a gravitational
smoothing length of $1.4$ comoving $\rm{kpc}$ for the dark matter particles. The mass of each dark matter particle is $5.1 \times 10^{6}\hMsun$ and the initial mass of gas particle is $9.4 \times 10^{5} \hMsun$.  The cosmological
parameters are based on \emph{Planck} \citep{2016A&A...594A..13P}.

\subsection{Shape calculation} \label{S:shapes}

The shapes of the stellar matter component in subhalos are modeled as
ellipsoids in three dimensions using 
the eigenvalues and eigenvectors of the reduced inertia tensor.

\begin{equation} \label{eq:redinertensor}
\widetilde{I}_{ij} = \frac{\sum_{n} m_{n}\frac{x_{n,i}x_{n,j}}{r_{n}^{2}}}{\sum_{n} m_{n}},
\end{equation}
where the summation is over particles indexed by $n$, and 
\begin{equation} \label{eq:rn2}
 r_{n}^{2} = \sum_{i}x_{n,i}^{2}\,.
\end{equation}
Here, $m_{n}$ is the mass of the $n^{\rm{th}}$ particle and $x_{n,i}$ is the position co-ordinate of the $n^{\rm{th}}$ particle for $0 \leq i \leq 2$ in $3\rm{D}$. We note that the effect of weighting instead with a star particle's luminosity instead of mass is investigated in \cite{2015MNRAS.448.3522T} and found to have negligible impact on the derived shape distributions. Similar to earlier studies, we only choose galaxies with a minimum of 1000 dark matter particles and 1000 star particles for resolved shapes of the dark matter and stellar component. Note that we impose this threshold on the number of particles for both the dark matter and stellar component. This is to ensure that while comparing the quantities based on the shapes of dark matter and stellar component, the shapes are well resolved for both. Further, since the minimum subhalo mass considered in this paper is atleast $10^{11} h^{-1}M_{\odot}$, we are already above the particle threshold for the number of dark matter particles. The reduced inertia tensor gives more weight to particles that are
closer to the center of the subhalo in question, which reduces the influence of loosely bound particles present in the outer regions of the subhalo. The shapes are determined iteratively using the reduced inertia tensor as detailed in \cite{2015MNRAS.448.3522T}. The axis ratios and eigenvectors In 3D, 
the eigenvectors of the inertia tensor are
${\hat{e}_{a}, \hat{e}_{b}, \hat{e}_{c}}$ with corresponding
eigenvalues 
$\lambda_{a} > \lambda_{b} > \lambda_{c}$. The eigenvectors represent
the principal axes of the ellipsoid, with the half-lengths of the principal
axes $(a,b,c)$ given by 
$(a,b,c) = (\sqrt{\lambda_{a}},\sqrt{\lambda_{b}},\sqrt{\lambda_{c}})$. The 3D
axis ratios are 
\begin{equation} \label{eq:axisratios}
q = \frac{b}{a}, \,\, s = \frac{c}{a}.
\end{equation}
 The
projected shapes are calculated by projecting the positions of the particles
onto the $XY$ plane of the simulation box and modeling the shapes as ellipses. In 2D, the eigenvectors are ${\hat{e}_{a}',\hat{e}_{b}'}$ with 
corresponding eigenvalues 
${\lambda_{a}' > \lambda_{b}'}$. The lengths of the semi-major and semi-minor axes
are $a' = \sqrt{\lambda_{a}'}$ and $b' = \sqrt{\lambda_{b}'}$ with axis
ratio $q_{2\rm{D}} = b'/a'$.

In the iterative method, we initially determine the principal axis lengths and the eigenvectors of the ellipsoid in the first iteration using all the particles of the given type in the subhalo. Then, keeping the enclosed volume constant, we rescale the lengths of the principal axes of the ellipsoid and recompute the shapes by discarding the particles outside the ellipsoidal volume. This process is repeated iteratively till a convergence criterion is reached. For convergence, we require that the fractional change in axis ratios in between successive iterations is below $1\%$.

\subsection{Shape of the satellite system} \label{S:shape_sat}

We define the shape of the satellite system for a given halo as the shape traced by the positions of the satellite galaxies of the halo. The shape is determined by the mass weighted inertia tensor given by

\begin{equation} \label{eq:shapecloud}
I_{ij}^{\rm{Sat}} = \frac{\sum_{n} m_{n} x_{n,i} x_{n,j}}{\sum _{n} m_{n}},
\end{equation}
where $m_{n}$ is mass of the $n^{\rm{th}}$ subhalo and $x_{n,i}$ refers to the satellite position. Similar to the shape calculation method, the eigenvector decomposition of the inertia tensor will give the major axes of the shape traced by the satellite system. This definition is similar to the one adopted by \cite{2016MNRAS.460.3772S}. Note that we use all the satellite subhalos in the given halo, without any threshold on the number of dark matter or star particles. However, the positions of more massive subhalos are given more weighting in the inertia tensor as seen from Equation~\ref{eq:shapecloud}. The alignment of the shape of the satellite system with that of the central galaxy will help us understand the distribution of satellite galaxies in the halo. A stronger alignment will indicate a tendency for the satellite galaxies to be distributed along the major axis of the central galaxy.

\section{Results} \label{S:results}

In this section, we discuss our findings on the anisotropic distribution of the satellite galaxies with the orientation of central galaxy as well as the orientation of satellite galaxies with respect to the central in MassiveBlack-II and IllustrisTNG simulations. 

\subsection{Alignment of central galaxy shape with shape of satellite system} \label{S:satsystem}
\begin{figure*} 
\begin{center}
\includegraphics[width=3.2in]{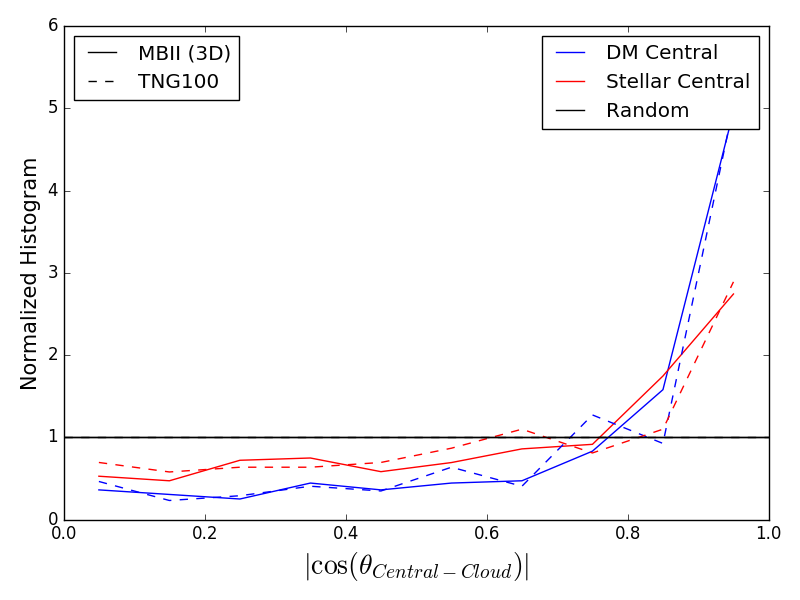}
\includegraphics[width=3.2in]{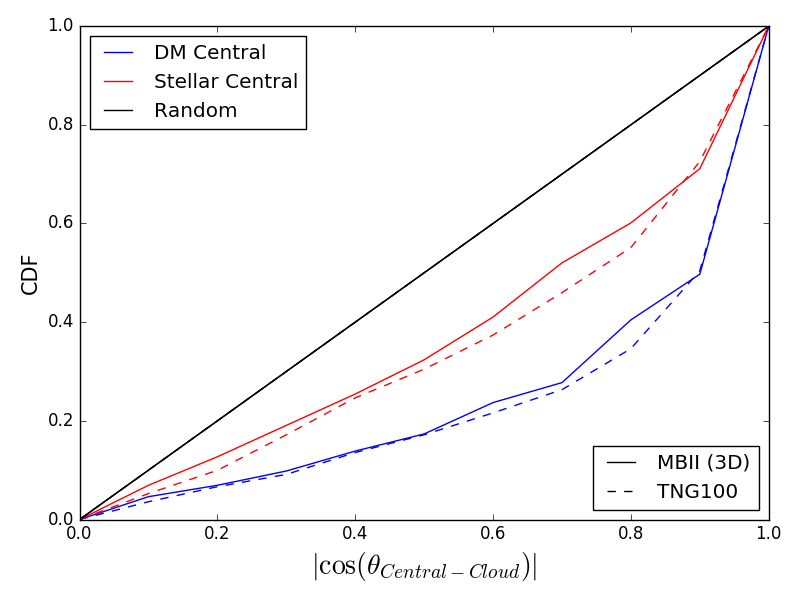}\\
\includegraphics[width=3.2in]{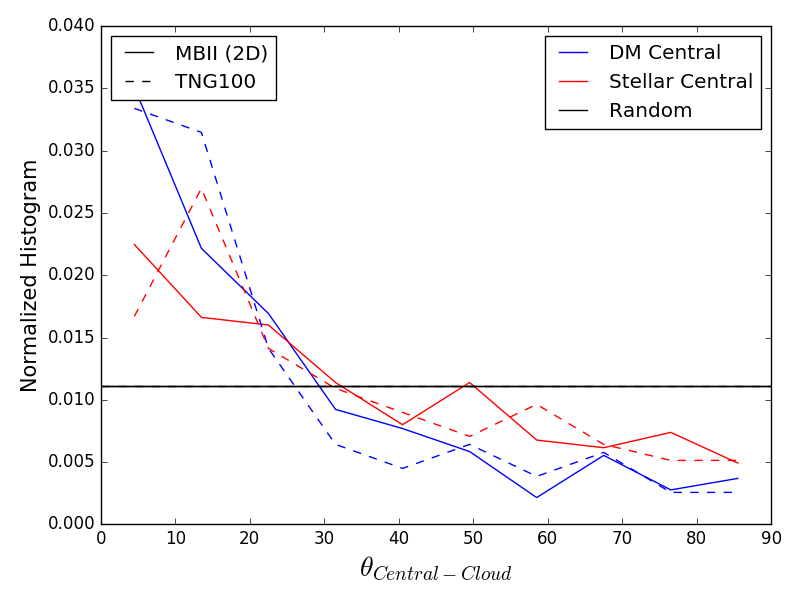}
\includegraphics[width=3.2in]{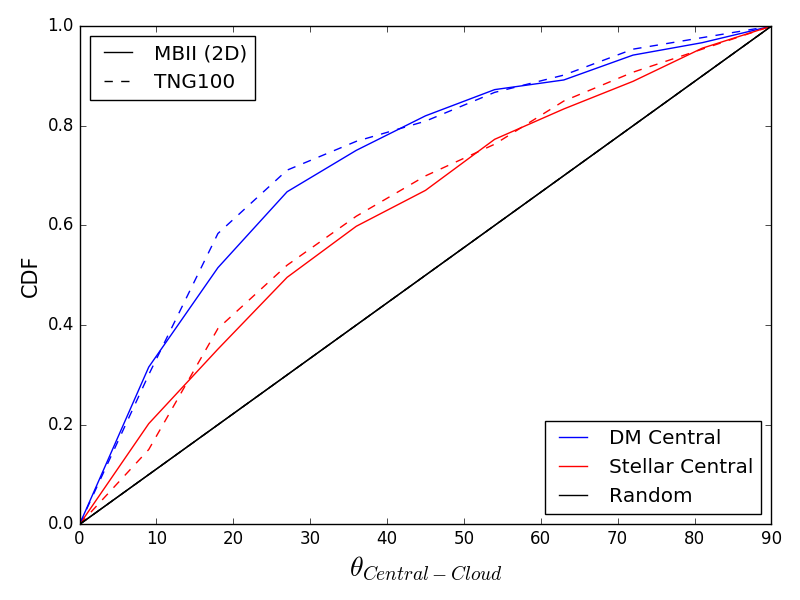}\\
\caption{\label{F:fig_cen_cloud} Alignment of the shape of the dark matter and stellar matter component of the central galaxies of halo mass ($>10^{13} h^{-1}M_{\odot}$) with the shape of the satellite system. {\textit{Top panel:}} $3\rm{D}$ alignments, {\textit{Bottom panel:}} $2\rm{D}$ alignments. The normalized distribution of alignment angles is shown in the {\textit{Left}} column, while the {\textit{Right}} column shows the cumulative distribution of the cosine of the alignment angles. The solid lines show the distributions for the MassiveBlack-II simulation and the dashed lines represent the Illustris-TNG simulation.}
\end{center}
\end{figure*}

\begin{figure*} 
\begin{center}
\includegraphics[width=3.2in]{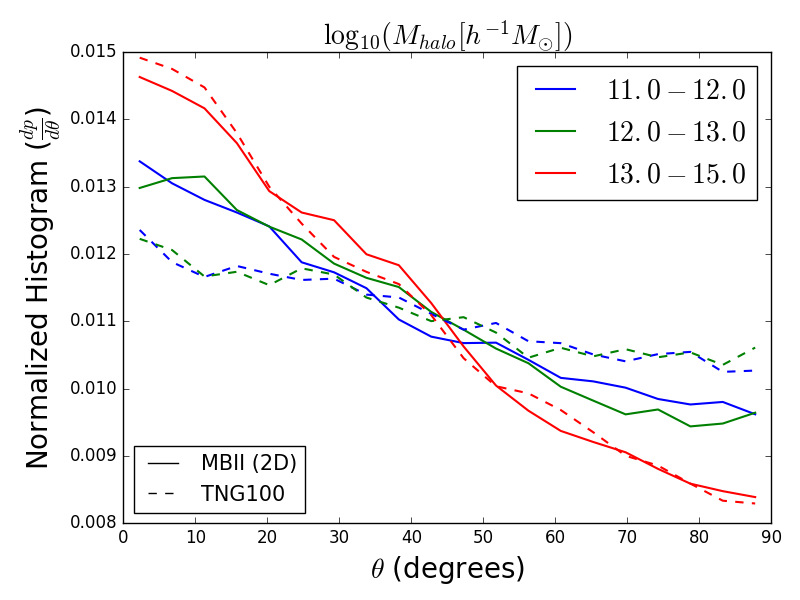}
\includegraphics[width=3.2in]{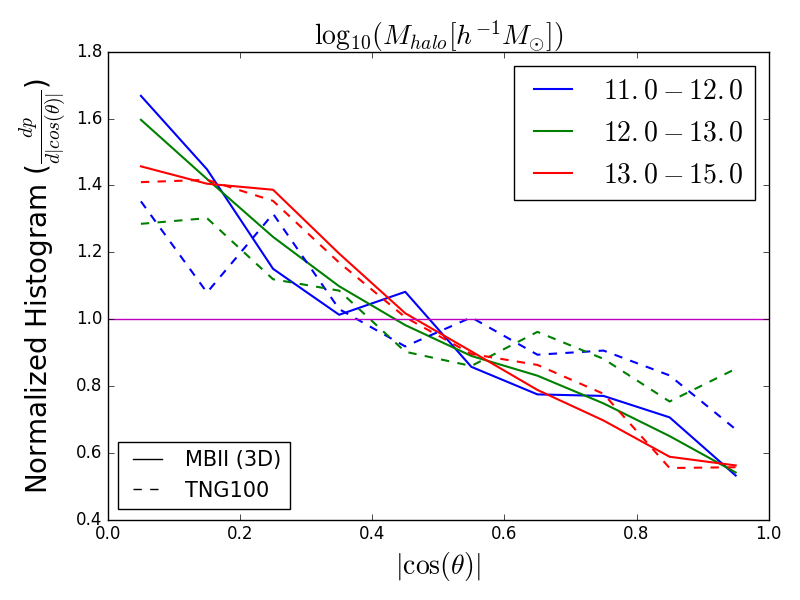}
\caption{\label{F:fig_cen_sathist} Normalized distribution of the alignment angles between the $2\rm{D}$ shape of the stellar matter component of the central galaxy and the location of satellite galaxies in MassiveBlack-II and Illustris-TNG (\textit{Left}). The (\textit{Right}) panel shows the distribution of the cosine of the angle between the minor axis of the stellar shape in $3\rm{D}$ and the location of the satellite galaxies (\textit{Right}). The histograms are shown for halos of mass, $10^{11-12} h^{-1}M_{\odot}$,  $10^{12-13} h^{-1}M_{\odot}$ and $10^{13-15} h^{-1}M_{\odot}$. The solid lines show the distributions for the MassiveBlack-II simulation and the dashed lines represent the Illustris-TNG simulation.}
\end{center}
\end{figure*}

We first explore the alignment between the shape of the dark matter and stellar matter component of the central galaxy with the shape traced by the satellite system within the halo. The shape of the satellite system is obtained using the positions of satellite galaxies within the halo as described in Section~\ref{S:shape_sat}. Here, we consider only halos with masses above $10^{13}h^{-1}M_{\odot}$ to allow for sufficient satellite galaxies at $z=0.06$. We use all the satellite galaxies of the halo without imposing any limit on the stellar particle number to trace the shape of the satellite system. However, we verified that the results are consistent when we apply a threshold on the number of star particles in the satellite galaxy (such as 50 star particles). 
In Figure~\ref{F:fig_cen_cloud}, we plot the distribution of the alignment angles between the major axis of the shape of the satellite system and the major axis of the shape of the central galaxy, determined by the dark matter and stellar matter component of the galaxy. The $3\rm{D}$ alignments are shown in the {\textit{Top panel}} of the figure, while the {\textit{Bottom panel}} shows the projected alignments. In the left hand side of the top panel, we show the normalized distribution of the $3\rm{D}$ alignments. We can see that the satellite system tends to be aligned with the shape of the central galaxy, when compared with that of a random distribution. Clearly, the satellite system has a higher degree of alignment with the dark matter component. This implies that on average, the satellite galaxies tend to be anisotropically distributed along the major axis of the shape of the central galaxy in comparison to the shape of stellar component. The mean alignment angles are $33.08 ^{\circ}  \pm 1.25 ^{\circ}$ and $44.19 ^{\circ} \pm 1.28 ^{\circ}$ respectively for the shape of dark matter component and stellar component respectively in Massive-Black II. In Illustris-TNG, the mean misalignment angles are $33.78 ^{\circ} \pm 1.85 ^{\circ}$ and $45.95 ^{\circ} \pm 1.89 ^{\circ}$. Compared with a uniform distribution this corresponds to $\Delta \theta$ of $\sim 12 ^{\circ}$ in MassiveBlackiII and TNG. 

Observationally, \cite{2016MNRAS.463..222H} studied the projected alignments of the central galaxy with the shape traced by the satellites in redMaPPer clusters and found a mean misalignment angle of $\sim 35 ^{\circ}$ and the alignment is found to be stronger in central galaxies of higher luminosity. Here, we do not investigate the mass dependence due to fewer satellites in halos of lower mass for resolving shapes. Further, it is found that satellites close to the central galaxies are more anisotropically distributed along the central major axis. We discuss the radial dependence of the anisotropic distribution in the subsections below. The right panels of Figure~\ref{F:fig_cen_cloud} show the cumulative distribution function (CDF) of the absolute value of the cosine of the alignment angle between the orientations of the shapes. It is to be noted that for random orientations in 3D, the cosine of the alignment angle follows a uniform distribution. Hence we show the CDF of $\cos (\theta _{Central-Cloud})$ in the top right panel of Figure~\ref{F:fig_cen_cloud}. We can clearly see that the degree of alignment is significant when compared to a random uniform distribution. Comparing the distributions in MassiveBlack-II and Illustris-TNG, we can see that the alignment trend is very similar and the cumulative distributions are consistent with each other. In the bottom panel of Figure~\ref{F:fig_cen_cloud}, we similarly show the alignment distributions in $2\rm{D}$. For 2D random orientations, the alignment angle follows a uniform distribution. Accordingly, the CDF of $\theta _{Central-Cloud}$ is plotted in the bottom right panel of Figure~\ref{F:fig_cen_cloud}. Further, the alignments are similar for both MB-II and Illustris-TNG, as in the case of $3\rm{D}$ alignments. Compared with a uniform distribution of alignment angles, the mean alignment of the satellite system in $2\rm{D}$ is stronger by $\Delta \theta \sim 12 ^{\circ}$ in both the simulations which is similar to the alignment strength in $3D$.  

In order to understand the mass dependent trend of the satellite distribution, we plot the histogram of orientation between the shape of the stellar component of the central galaxy and the location of satellite galaxies in Figure~\ref{F:fig_cen_sathist} for halos of mass range, $10^{11-12} h^{-1}M_{\odot}$,  $10^{12-13} h^{-1}M_{\odot}$ and $10^{13-15} h^{-1}M_{\odot}$. From the figure, we can observe a mass dependent trend in the distribution of satellites in the MassiveBlack-II simulation. The satellite galaxies tend to be distributed along the major axis of the central galaxy with a higher degree of alignment in high mass halos. From previous studies based on MassiveBlack-II \citep{2015MNRAS.448.3522T}, it is known that the stellar shape is more misaligned with the shape of dark matter halo in halos of lower mass. Since, the satellite system is more aligned with the shape of dark matter halo as discussed earlier, we expect the satellite distribution to be more anisotropic in high mass halos, due to the larger alignment of the stellar shape. The trend is similar in the Illustris-TNG simulation as well. From the figure, we can also observe that the satellite distribution is more anisotropic in the MassiveBlack-II simulation for the lower halo mass bins. In a previous study \citep{2016MNRAS.462.2668T}, the stellar component of galaxies in Illustris simulation is found to be more misaligned with the dark matter component when compared with the MassiveBlack-II simulation. These differences in galaxy alignments are likely due to the different baryonic feedback models adopted in the two simulations. Since the feedback model adopted in Illustris-TNG is more similar to that of the Illustris simulation, the stronger anisotropic distribution of satellite galaxies in MassiveBlack-II is consistent with \cite{2016MNRAS.462.2668T} results.  \cite{2015arXiv151200400W} found that the satellite galaxies tend to be distributed along the plane of the central galaxy perpendicular to the direction of the minor axis. We similarly plot the histogram of the orientation of the minor axis of the central shape with the location of the satellite positions. As seen from the histogram of the cosine of the angle in the right panel of Figure~\ref{F:fig_cen_sathist}, we see that there is an excess probability for the cosine of the angle to be $0$, which indicates that the satellites are distributed along the plane of the galaxy. We do not find a significant mass dependent trend here. 

\subsection{Distribution of satellite galaxies with orientation of central galaxy: dependence on central galaxy properties}

\begin{figure*} 
\begin{center}
\includegraphics[width=3.2in]{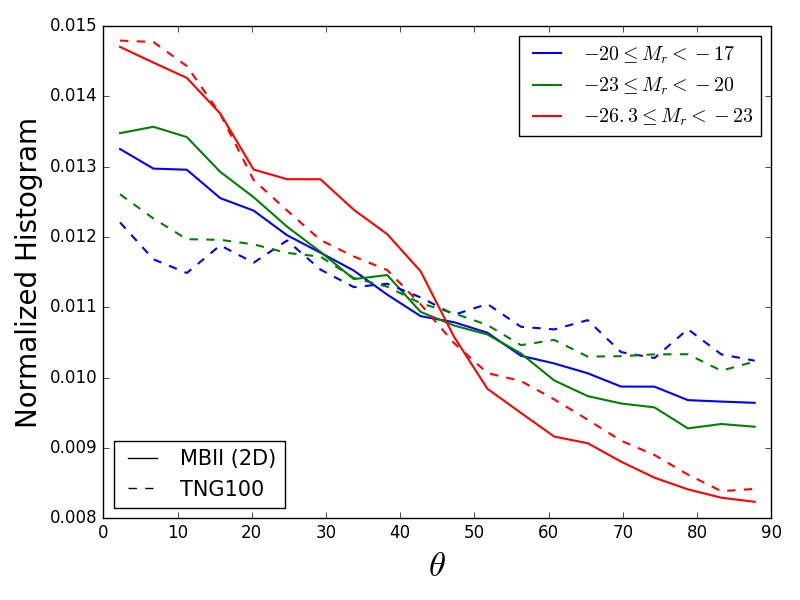}
\includegraphics[width=3.2in]{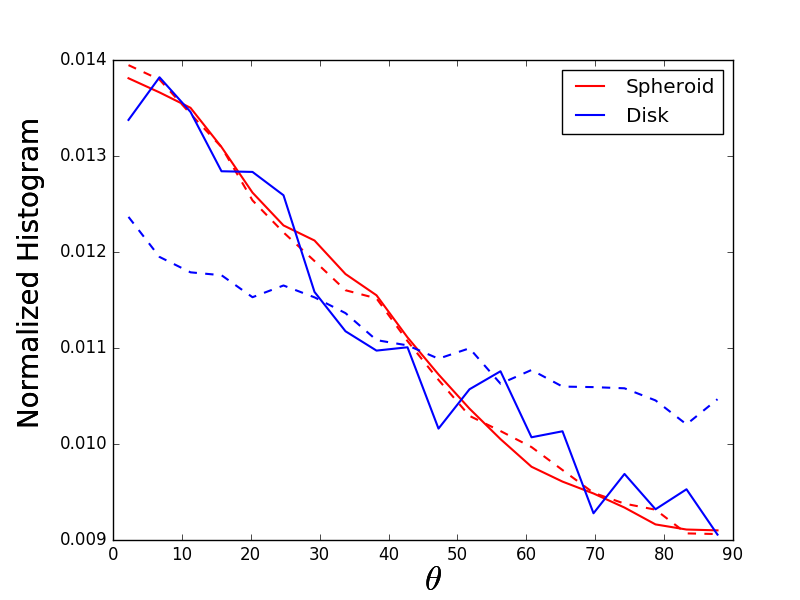}\\
\includegraphics[width=3.2in]{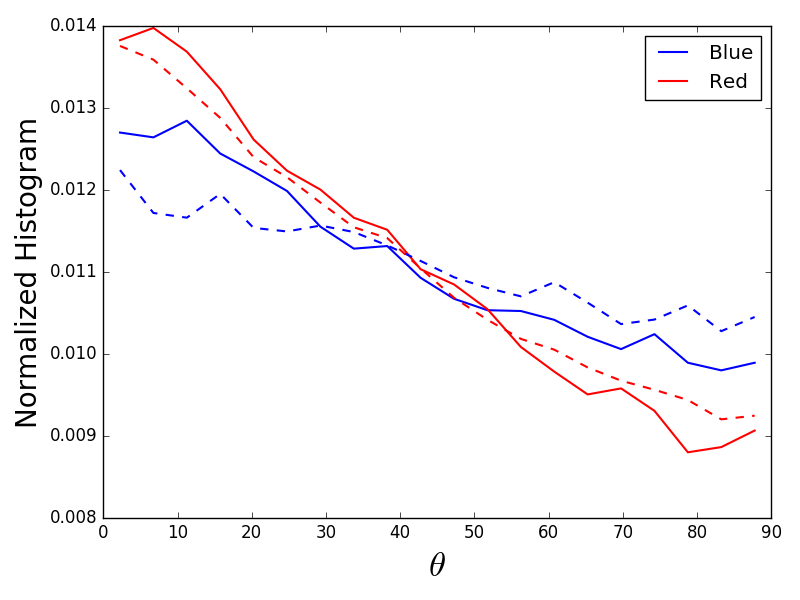}
\includegraphics[width=3.2in]{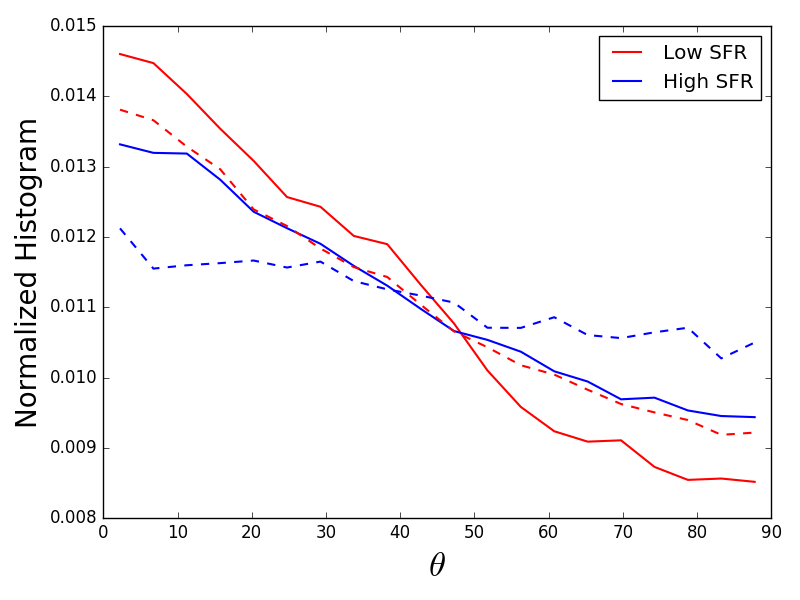}\\
\caption{\label{F:fig_satdist_prop} Normalized distribution of the alignment angles between the $2\rm{D}$ shape of the stellar matter component of the central galaxy and the location of satellite galaxies in MassiveBlack-II and Illustris-TNG simulations based on the central galaxy properties: Luminosity({\textit{top left}}), morphology({\textit{top right}}), Color({\textit{bottom left}}), star formation rate({\textit{bottom right}}). Here, the solid lines show the distributions for the MassiveBlack-II simulation and the dashed lines represent the Illustris-TNG simulation.}
\end{center}
\end{figure*}

Here, we consider the dependence of the anisotropic distribution of satellite galaxies on properties of the central galaxy. In Figure~\ref{F:fig_satdist_prop}, we plot the distribution of satellite galaxies based on the properties, such as luminosity, morphological type, color and star formation rate. Here, the luminosity of the galaxy is based on the SDSS $r$-band absolute magnitude in the rest frame ($M_{r}$). We define three luminosity bins, with $-26.3 \leq M_{r} < -23$, $-23 \leq M_{r} < -20$ and $-20 \leq M_{r} < -17$. The galaxy colors ($g-r$) are defined as the difference of the rest-frame absolute magnitudes in the SDSS $g$-band and SDSS $r$-band. In this study, we chose the median of the $g-r$ colors to split the galaxy sample into red and blue galaxies. The galaxies with low and high SFR are similarly defined based on the median of the galaxy star formation rates in the sample. 

The classification of galaxies in MassiveBlack-II into disks and spheroids is done based on a dynamical bulge-disc decomposition \citep[as discussed in][]{2016MNRAS.462.2668T} where all the galaxies with bulge-to-total ratio less than $0.7$ are classified as disks. The disk galaxies in Illustris-TNG are similarly defined. 

We can see that the luminosity trend is similar to that of the halo mass, with more satellites distributed along the major axis of galaxies with larger luminosity. \cite{2016MNRAS.460.3772S} found a dependence in the alignment of satellite distribution based on whether the central galaxy is disc or spheroid. To compare with their results, we similarly split the central galaxies in our sample into discs and spheroids.  However, we do not find a significant difference in the satellite distribution in MassiveBlack-II. As seen from the figure, the distribution of satellites is similar for both discs and spheroid galaxies. However, we find a color-dependent trend in the satellite distribution, with the red central galaxies showing a larger anisotropic distribution of the satellites along their major axes. The Illustris-TNG simulation results show a stronger tendency for the satellites in red central galaxy and spheroid galaxies to be distributed along the major axis. By comparing the satellite distribution in central galaxies split based on their star formation rate (SFR), we find a slightly higher tendency for the satellites in centrals of low SFR to be distributed along the major axis of centrals in both MassiveBlack-II and Illustris-TNG. We also find that the coplanar distribution of the satellite galaxies is stronger in spheroids, red centrals and galaxies with low SFR.

\subsection{Small-scale alignments : Correlation functions and comparison with observational measurements} \label{S:1hcorr}
In this Section, we consider the radial dependence of the correlation of a central galaxy shape with the location of satellites and the shape of the satellite galaxy with the location of its host halo. Accordingly, we calculated the small-scale shape-density correlation functions (within the halo, corresponding to the $1$-halo term) in $3\rm{D}$. We also consider projected radial ellipticities of the galaxy shapes for comparison with observational measurements. Throughout the rest of the paper, the error bars shown on the correlation functions correspond to poisson error bars for both the $3\rm{D}$ and projected statistics.

\begin{figure*} 
\begin{center}
  \includegraphics[width=3.2in]{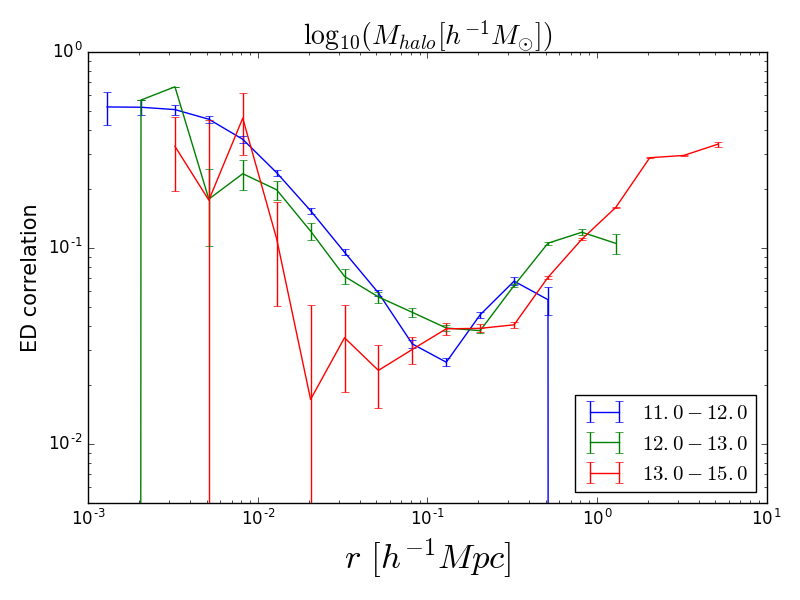}
  \includegraphics[width=3.2in]{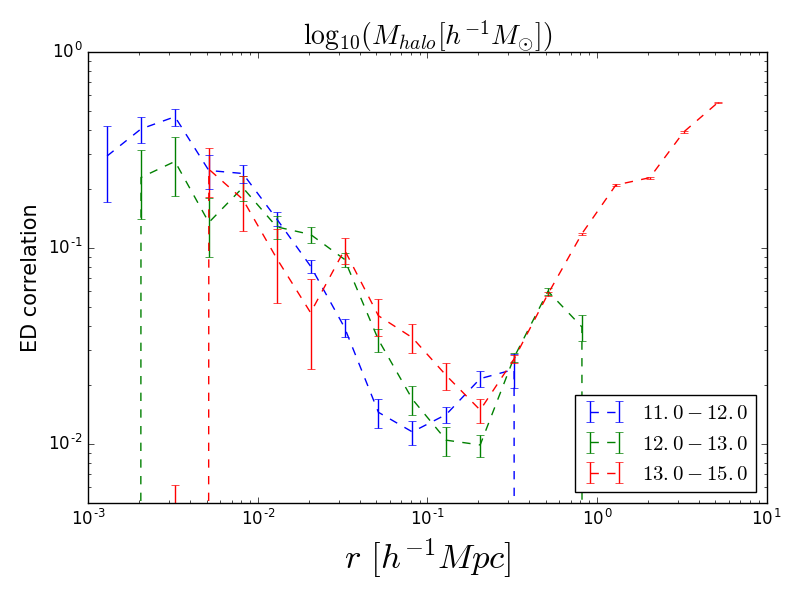}
\caption{\label{F:fig_ed_cen_sat} Ellipticity-direction (ED) correlation function of the $3\rm{D}$ shape of the central galaxy with the location of the satellite galaxies within the given halo in MassiveBlack-II ({\textit{Left}}) and Illustris-TNG ({\textit{Right}}) simulations. The correlation functions are plotted in halos of mass bins, $10^{11-12}h^{-1}M_{\odot}$, $10^{12-13}h^{-1}M_{\odot}$ and $10^{13-15}h^{-1}M_{\odot}$ at redshift, $z=0.06$. Here, we cross-correlate with the positions of all subhalos in the halo without a mass threshold. The non-negligible signal on a $\mathrm{kpc}$ scale is due to density peaks identified as subhalos. If we include a mass threshold in the density sample (satellite gaalxies), we do not see a signal below $\sim 5 h^{-1}\mathrm{kpc}$.}
\end{center}
\end{figure*}

In order to understand the radial dependence of the distribution of satellite galaxies along the major axis of the central galaxy, we consider the Ellipticity-Density (ED) correlation function for the shape of the stellar component of central galaxies in Figure~\ref{F:fig_ed_cen_sat}. For a galaxy with orientation of major axis, $\hat{e}(\textbf{r})$ and density tracers located at a distance $r$ in the direction of the unit vector, $\hat{r}$, the ED correlation is given by the mean,

\begin{equation} \label{eq:ed}
  \omega(r) = \langle |\hat{e(\textbf{r})} \cdot \hat{r}| ^{2} \rangle - \frac{1}{3}. 
\end{equation}
Here the shape of the central galaxy is cross-correlated with the satellite positions only within the halo corresponding to the central galaxy. We note that for random orientations in $3D$, the value of $\omega (r)$ is 0. A positive value of $\omega (r)$ indicates that the density tracers are distributed more along the major axis. For perfect alignments, i.e, when the direction of major axis points along the density tracers, $\omega (r)$ is maximum and has a value of $\frac{2}{3}$. For the density tracers located in the plane normal to the orientation of the major axis, the value of $\omega (r)$ is $\frac{-1}{3}$.

  The left panel of Figure~\ref{F:fig_ed_cen_sat} shows the ED correlation function in the MassiveBlack-II simulation, while the right panel corresponds to the results in IllustrisTNG simulation. The correlation function is plotted for the halos within the mass bins, $10^{11-12}h^{-1}M_{\odot}$, $10^{12-13}h^{-1}M_{\odot}$, and $10^{13-15}h^{-1}M_{\odot}$ by cross-correlating the shape of a central galaxy with the location of the satellite galaxies inside the halo. As noted earlier in this section, we plot the Poisson error bars for the correlation functions. At very small scales ($< 0.1 h^{-1}\rm{Mpc}$), we can see that the correlation function shows a decreasing trend with distance from the galaxy center. This indicates that the satellite galaxies close to the central galaxy are distributed along the major axis of the stellar shape and tend to be more symmetrically distributed as the distance from the central galaxy increases. However at larger distances close to the halo boundary ($ \sim 0.1-1 h^{-1}\rm{Mpc}$), we can clearly see an increase in the correlation function. This is because at larger distances, the satellites tend to be more anisotropically distributed along the major axis of the shape of the dark matter component of the galaxy with increasing distance. Since, the stellar shape is in turn correlated with the shape of the dark matter component, we also see a similar radial dependence of the distribution of the satellites w.r.t central galaxy on these scales (with smaller amplitude). 

\begin{figure} 
\begin{center}
\includegraphics[width=3.2in]{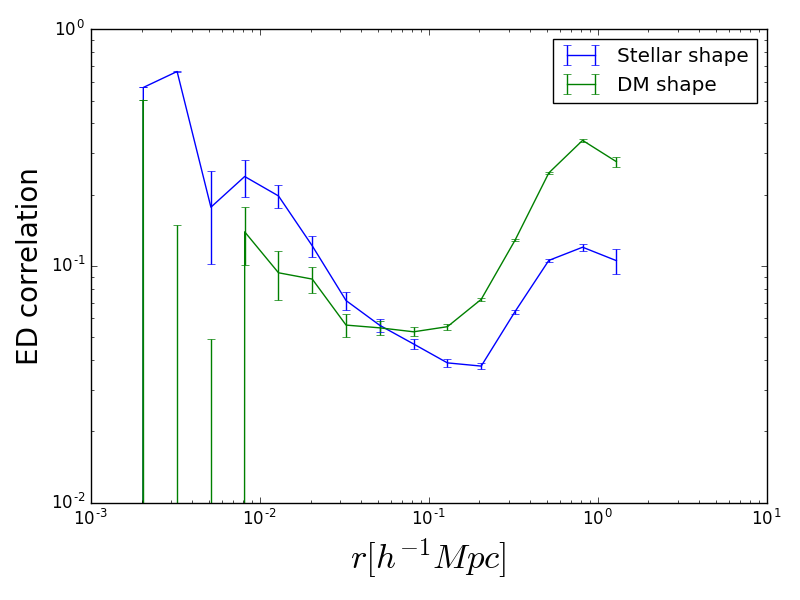}
\caption{\label{F:fig_ed_censhape_tracers}
  ED correlation for the shape of the dark matter and stellar component of the central galaxies in the MassiveBlack-II simulation with the density distribution traced by the positions of satellites 
  within the corresponding halo. The correlation function is for the halos of mass, $10^{12-13}h^{-1}M_{\odot}$ at $z=0.06$.
  }
\end{center}
\end{figure}

\begin{figure*} 
\begin{center}
  \includegraphics[width=3.2in]{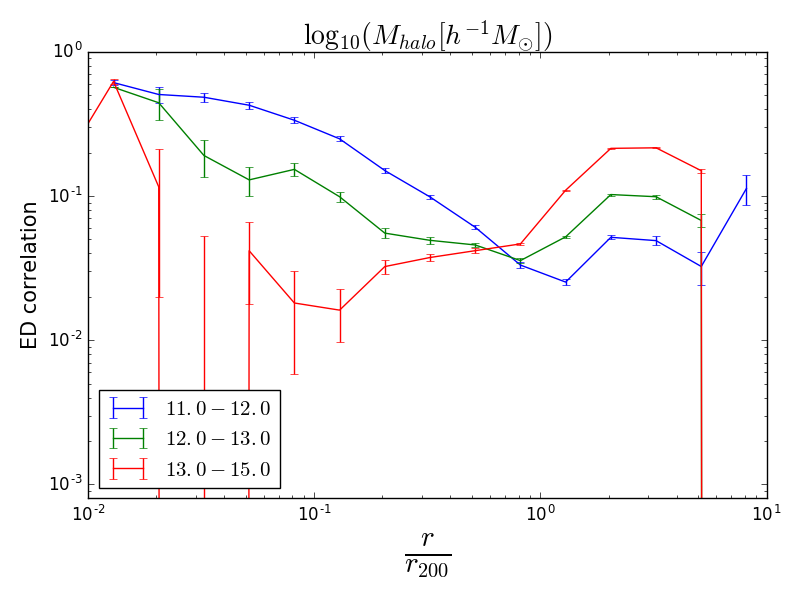}
  \includegraphics[width=3.2in]{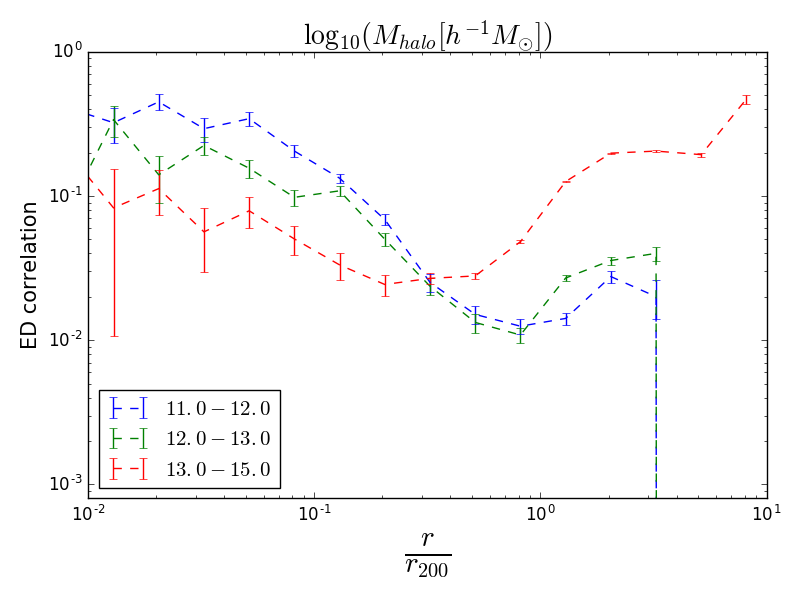}
\caption{\label{F:fig_ed_cen_sat_r200bins1} Ellipticity-direction (ED) correlation function of the $3\rm{D}$ shape of the central galaxy with the location of the satellite galaxies within the given halo in MassiveBlack-II ({\textit{Left}}) and Illustris-TNG ({\textit{Right}}) simulations where the radial binning is in $\frac{r}{r_{200}}$.
}
\end{center}
\end{figure*}

  This is further illustrated in Figure~\ref{F:fig_ed_censhape_tracers}, where the shape of the dark matter and stellar matter component of the galaxy is correlated with the location of satellites 
  within the halo. At small scales, we can see that the amplitude of ED correlation function is higher for the shape of the stellar component while on large scales, it is larger for the shape of dark matter component. This indicates that on small scales, the satellites are distributed anisotropically along the major axis of the shape of the stellar component, while on large scales, they are distributed along the shape of the dark matter component. On intermediate scales, the satellite distribution tends to be uncorrelated with the shape of stellar component and hence, we observe a decreasing correlation function. The increase in correlation on large scales for the shape of stellar component is due to the correlation of the stellar shape with that of the dark matter component.

  Comparing the ED correlation function in MassiveBlack-II and IllustrisTNG in Figure~\ref{F:fig_ed_cen_sat}, we find that the radial dependence (a decreasing correlation function, followed by an increase at scales closer to the limit of $1$-halo regime) is similar in both simulations. We can also find a mass-dependence of the correlation functions with a higher amplitude at larger scales in halos of high-mass. There is also a mass-dependence in the scale at which the correlation function shows a tilt from the decreasing trend. However, at intermediate scales ($\sim 0.01 - 0.1 h^{-1} M_{\odot}$), we see that the correlation function is larger in low mass halos in the MassiveBlack-II simulation, while in the IllustrisTNG simulation, it is larger in high-mass halos. This means that the mass-dependence of the amplitude of the correlation function in the intermediate range varies between the two simulations. It's higher in low mass halos of MassiveBlack-II, whereas in IllustrisTNG, the amplitude is larger for high mass halos. \cite{2017MNRAS.467.4131V} used the GAMA galaxy groups from weak lensing and found that the satellites in the outer halo trace the orientation of the dark matter halo. Similarly, consistent with our findings, the satellite distribution is found to be anisotropic with respect to the major axis of the BCG and the signal decreases with the distance from the center of the galaxy.
  
  In Figure ~\ref{F:fig_ed_cen_sat_r200bins1}, we plot the ED correlation function with bins in $r/r_{200}$. This is to understand the dependence of the scale at which we find a transition of the signal from a decreasing trend to an increasing one on large scales and if it is related to the average $r_{200}$ values of the halo sample. From the figure, we can see that there is no clear dependence of this transition scale on $r_{200}$. This means that the scale at which the turnover from decreasing correlation to an increasing trend occurs is not correlated with the $r_{200}$ values of the halo sample. Instead, we find a clear mass dependence in the amplitude of the signal at small and large scales. The satellite distribution on smaller scales is more anisotropic along the major axis of the stellar shape in low mass halos as seen by the larger amplitude of ED correlation. However, as known from previous studies \citep{2015MNRAS.448.3522T}, the stellar shape in low mass galaxies is more misaligned with the shape of the dark matter subhalo, which leads to a lower correlation with satellite distribution on larger scales, which tends to be more correlated with the shape of the dark matter halo.

\begin{figure*} 
\begin{center}
  \includegraphics[width=3.2in]{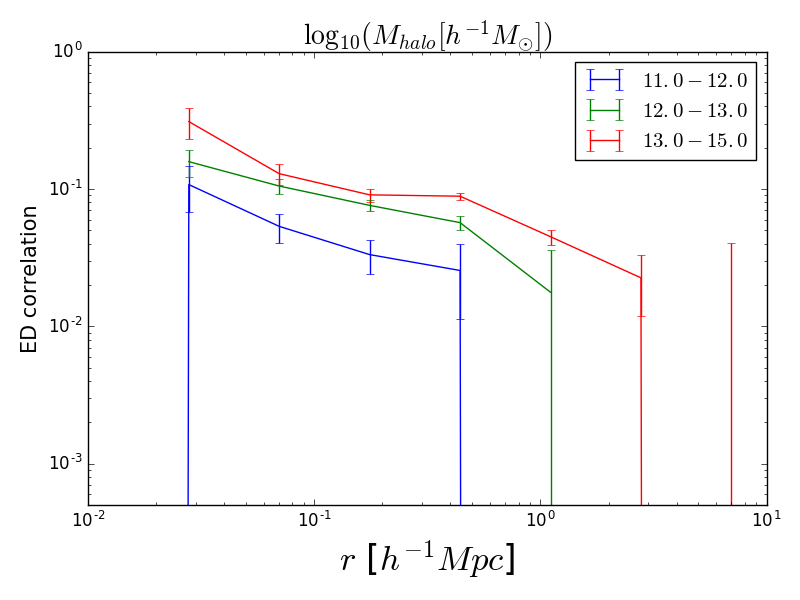}
  \includegraphics[width=3.2in]{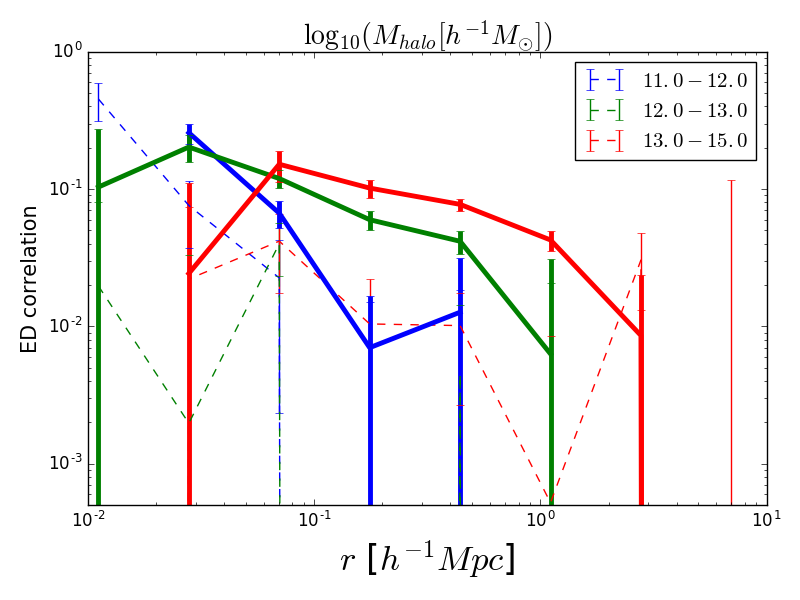}\\
\caption{\label{F:fig_ed_sat_shapes} Ellipticity-direction (ED) correlation function of the $3D$ shape of the satellite galaxies with the location of the central galaxy within the given halo in MassiveBlack-II ({\textit{Left}}) and Illustris-TNG ({\textit{Right}}) simulations.  
The thick solid lines in the {\textit{Right}} panel of the plot show the correlation function in Illustris simulation. (Note that we do not refer to the Illustris-TNG simulation here.)}
\end{center}
\end{figure*}

Similarly, we can also understand the radial dependence of the orientation of the shapes of the satellite galaxies with respect to the location of central galaxy. In the top panel of Figure~\ref{F:fig_ed_sat_shapes}, we plot the ED correlation of the shape of satellite galaxies in a halo of given mass with respect to the location of the central galaxy. From the figure, we can see that the correlation function decreases as the radial separation increases from the center of the halo. This indicates that the satellite shapes closer to the halo center have a stronger alignment towards the center. We can also see that the amplitude of alignment shows a mass dependent trend with a higher alignment in more massive halos. 

Our results are consistent with observational measurements of \cite{2018MNRAS.474.4772H} in redMaPPer clusters. \cite{2018MNRAS.474.4772H} found that the alignments of satellite ellipticites with central galaxies decreases with the radial distance from central galaxy and is stronger in more luminous satellites. However, for Illustris-TNG we only find a decreasing trend with radial distance for the lowest mass bin ($10^{11-12}h^{-1}M_{\odot}$). For higher mass bins, the correlation is not significant and does not seem to exhibit a radial dependence. It is possible that this feature in Illustris-TNG simulation is due to baryonic feedback \citep{{2017ApJ...834..169T},{2019arXiv190811665S}}. In order to understand this further, we also plot the alignments of satellites in Illustris simulation. These are shown by the dashed lines in the plot. Here, we can see that the radial dependence and halo mass dependence of satellite alignments is similar to what is observed in the MassiveBlack-II simulation. Since the numerical hydrodynamic scheme is similar in Illustris and Illustris-TNG, the differences can be attributed to baryonic feedback models.

\begin{figure*} 
\begin{center}
\includegraphics[width=3.2in]{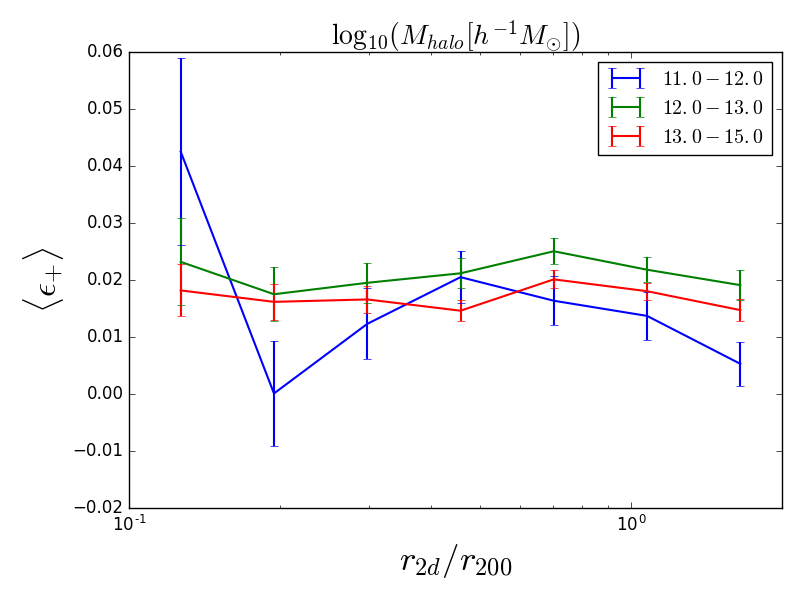}
\includegraphics[width=3.2in]{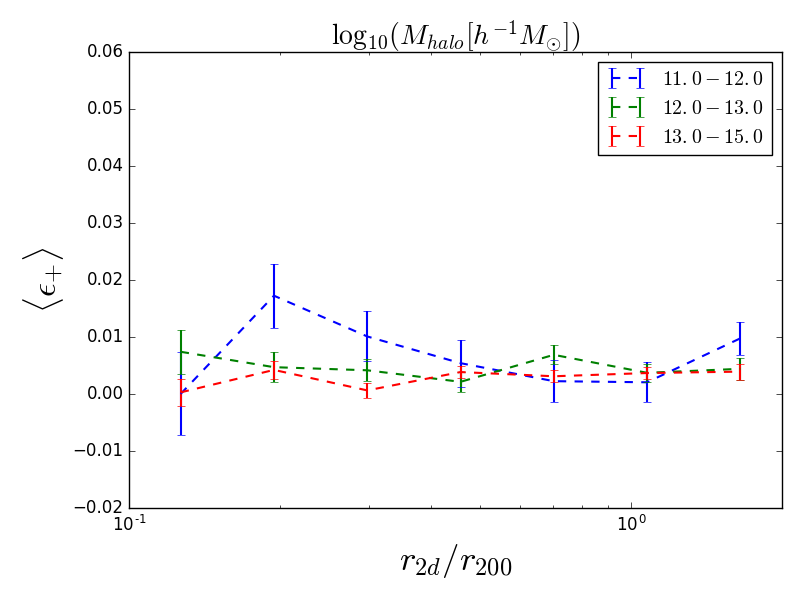}
\caption{\label{F:fig_avg_ellip} The mean of the radial component of the projected ellipticities of the shapes of the stellar component of the satellite galaxies with respect to the host central galaxy. Here, the binning is in $\frac{r_{2d}}{r_{200}}$, the ratio of the projected distance of the satellite to the central galaxy, $r_{2d}$ and the $r_{200}$ of the central galaxy. The mean radial ellipticities are shown for the MassiveBlack-II ({\textit{Left}}) and Illustris-TNG ({\textit{Right}}) simulations in halo mass bins of $10^{11-12}h^{-1}M_{\odot}$, $10^{12-13}h^{-1}M_{\odot}$ and $10^{13-15}h^{-1}M_{\odot}$}
\end{center}
\end{figure*}

We compare the measurements of the radial alignment of satellite galaxies in the simulation with recent observational measurements. \cite{2019arXiv190500370G} measured the projected ellipticities of the shape of satellite galaxies with respect to the location of the central galaxy in GAMA+KiDS. \cite{2019arXiv190500370G} measured the average ellipticity in radial bins and found that the average radial ellipticity decreases with radial distance. In order to compare with these measurements, we plot the mean of the radial component of the projected ellipticity, $\langle \epsilon _{+} \rangle$ in Figure~\ref{F:fig_avg_ellip}. The radial component of the ellipticity is calculated by

\begin{equation} \label{eq:ellip}
  \epsilon _{+} = \frac{1-q}{1+q} \cos [2 \theta]
\end{equation}
where, $\theta$ is the orientation of the major axis of the shape of satellite galaxy with respect to the halo center. We note that while computing the projected quantities, the projected ellipticities can be measured along the $\rm{XY}$, $\rm{YZ}$ or $\rm{ZX}$ planes. Here, for each galaxy, we measured the projected quantities along all the three planes in order to improve the correlation function statistics.

In Figure~\ref{F:fig_avg_ellip}, we plot the mean radial ellipticity in halos of different mass bins. In both IllustrisTNG and MB-II simulations, we do not find a significant mass dependent trend in our measurements. While the satellite alignments in $3$D show a radial dependence in the MassiveBlack-II simulation, the radial dependence is diluted in the alignment of the projected shapes with respect to the central galaxy. We further discuss the effects of projected shapes and projected distances on the radial alignment signal in Appendix~\ref{app:projection}. However, the amplitude of the signal is roughly of the same order of magnitude as the observational measurements \footnote{We note that there are no significant quantitative differences in the results when adopting the alternative definition for the ellipticity, $\epsilon = (1/2R)[(1- q^2)/(1 + q^2)]$, where $R = 1 - e_{\rm rms}^{2}$ is the shear responsivity factor, with $e_{\rm rms}$ being the rms ellipticity per component of the galaxy shape sample}. It is to be noted that an exact comparison with observational measurements is difficult due to differences in shape measurement methods adopted and selection effects in observational data. 

\begin{figure*} 
\begin{center}
  \includegraphics[width=3.2in]{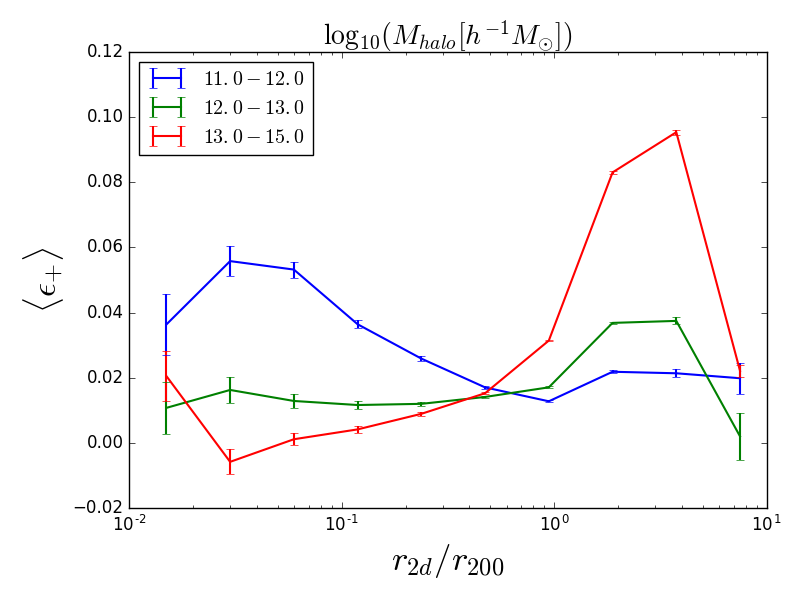}
  \includegraphics[width=3.2in]{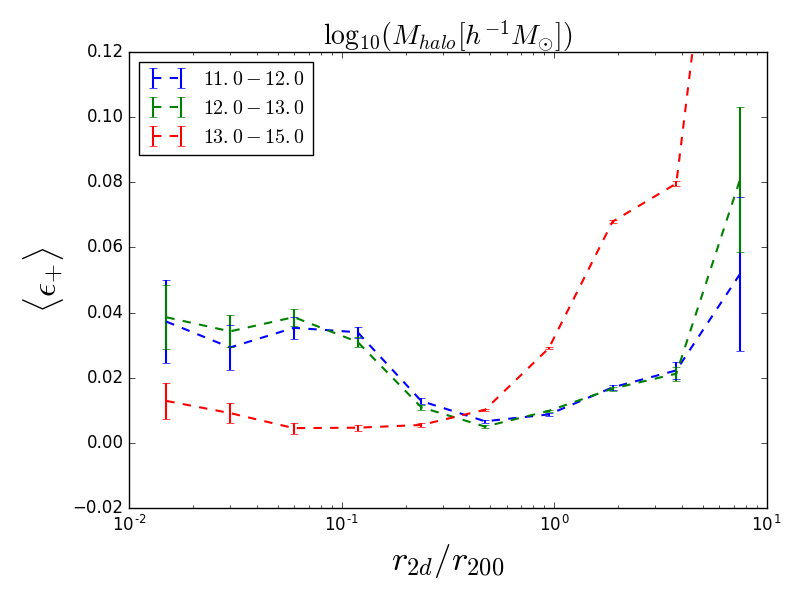}
\caption{\label{F:fig_ellip_cen} Mean radial ellipticity of the projected shape of the central galaxy with the location of the satellite galaxies within the given halo in MassiveBlack-II ({\textit{Left}}) and Illustris-TNG ({\textit{Right}}) simulations. Here, the binning is in $\frac{r_{2{\rm d}}}{r_{200}}$, the ratio of the projected distance of the satellite to the central galaxy, $r_{2d}$ and the $r_{200}$ of the central galaxy.
}
\end{center}
\end{figure*}

We also compare the radial dependence of the projected ellipticity component of the central galaxy with respect to the location of the satellite galaxies in Figure~\ref{F:fig_ellip_cen}. The observational measurements found a decreasing trend with the radial distance. In MassiveBlack-II simulation, we notice a decreasing trend only in the lowest mass bin, $10^{11-12}h^{-1}M_{\odot}$. The radial dependence is not significant in the halo mass bin, $10^{12-13} h^{-1}M_{\odot}$, while the highest mass bin exhibits an increase in the mean projected ellipticity with distance from center. This can be compared with the radial dependence trend for the $3$D shapes shown in Figure~\ref{F:fig_ed_cen_sat}. Similar to the ED correlation function in Figure~\ref{F:fig_ed_cen_sat}, the mean ellipticity for the highest mass bin increases at large scales due to the anisotropic distribution of the satellites along the major axis of the dark matter shape. For the lower mass bins, the signal at large scales is diluted due to projection effects. We find a similar dependence in the Illustris-TNG simulation results shown in the right panel. The small-scale correlation functions in the larger volume Illustris-TNG 300 simulation and a detailed comparison of the different correlation functions in MassiveBlack-II, Illustris-TNG and Illustris-TNG300 simulations within the same halo mass bins is discussed in Appendix~\ref{app:tng300}. We find that the mass and radial dependent trend of the correlation functions is similar across all the simulations. However, the amplitude of alignment is larger in MassiveBlack-II.

\section{Conclusions} \label{S:conclusions}

In this paper, we study the anisotropic distribution of satellite galaxies along the orientation of the major axis of the stellar component of the central galaxy in MassiveBlack-II \citep{2015MNRAS.450.1349K} and IllustrisTNG \citep{2019ComAC...6....2N} simulations. The mitigation of intrinsic alignments in upcoming surveys such as LSST, \emph{Euclid} and \emph{WFIRST} requires a small-scale modelling of the galaxy alignments. Currently, halo model approaches for intrinsic alignments \citep{Schneider2010} assume that the satellite galaxies are distributed spherically symmetric around the central galaxy and their orientation is pointed towards the location of central galaxy. We explore the deviations of galaxy alignments from the assumptions of halo model in realistic simulations of galaxy formation. We also compare our results with the measurements from observations and findings in other comparable simulations such as the EAGLE \citep{2015MNRAS.446..521S} and Horizon-AGN \citep{2014MNRAS.444.1453D}.

We first explore the alignment of the central galaxy shape with the shape of the satellite system traced by the positions of the satellite galaxies within the halo. In halos of mass greater than $10^{13}h^{-1}M_{\odot}$ at $z=0.06$, we find that the satellite system is well aligned with the shape of central galaxy when compared to a random spherically symmetric distribution. This indicates that the satellites are distributed anisotropically along the major axis of the central galaxy, instead of a symmetric distribution. This is consistent with the findings in EAGLE simulation by \cite{2016MNRAS.460.3772S}. \cite{2015arXiv151200400W} also found similar alignment between the orientations of the minor axes of the galaxy in Horizon-AGN. We compared the distribution of satellite galaxies along the major axis of the central galaxy based on properties such as halo mass, luminosity, morphology, color and star formation rate. We found that the satellites tend to be more strongly distributed along the major axis in galaxies of larger halo mass. This trend is similar in both MassiveBlack-II and IllustrisTNG simulations. Similarly, the distribution of satellites is stronger in galaxies with higher luminosity. We also find that the satellite distribution is more anisotropic around central galaxies of lower star formation rate, spheroid morphology and red centrals.

We also explored the radial dependence of the satellite distribution and orientation of the satellite galaxy with the respect to the orientation and location respectively, of the major axis of the central galaxy of the host halo. Looking at the radial alignments in $3\rm{D}$, we find that at small scales, the satellites are distributed along the major axis of the central galaxy and the correlation decreases as the distance from the central galaxy increases. However at distances reaching the boundary of the halo, we find that the correlation increases. This is due to the satellites on the outer regions of the halo being more anisotropically distributed along the orientation of the shape of the dark matter component of the galaxy, which is correlated with the shape of the stellar component.

We studied the radial dependence of the shape of the satellite galaxies with respect to the location of central galaxy. in MassiveBlack-II, we find that the galaxies at smaller distances from the central tend to point more strongly towards the location of central galaxy. The correlation decreases as distance from the center increases. We also find a similar trend in the Illustris simulation. However, we do not find a significant radial trend of satellite orientation in the IllustrisTNG simulation. The different trends in Illustris and IllustrisTNG simulation points to the effects of variations in baryonic feedback on satellite galaxy alignments. Finally, we compared the simulation measurements of projected correlation functions on small scales with the observational measurements from the GAMA+KiDS galaxies \cite{2019arXiv190500370G}. From observations, \cite{2019arXiv190500370G} found a radially decreasing trend of the satellite ellipticity with respect to the location of central galaxy. However, in MassiveBlack-II and IllustrisTNG, we do not find a radial trend in the mean satellite ellipticity. However, the amplitude of the mean radial ellipticity is consistent with the observational measurements. We note that the projection effects decreases the radial dependence of the satellite alignments in simulations and leads to a scale-independent radial trend.

Finally, comparing the results in MassiveBlack-II and Illustris-TNG simulations, we find that the various statistics quantifying satellite anisotropy and radial alignment are qualitatively consistent across the two simulations. In particular, we find agreement in the radial scaling of the satellite distribution with respect to the shape of the central galaxy in both MBII and TNG. The satellite anisotropy is found to be higher in central galaxies of larger halo mass, luminosity, low SFR and red galaxies. The qualitative agreement in these quantities is present for both 3D and the projected statistics measured for observational comparisons. Hence, this indicates that these numerical predictions are robust to the feedback implementations of galaxy formation physics. However, comparing the alignment of $3D$ satellite shape with central position, we do not find a significant mass dependent amplitude and a radially decreasing trend seen in MassiveBlack-II and Illustris. The lack of consitency in TNG and Illustris results indicates an effect of baryonic feedback prescriptions. Similarly, the anisotropic distribution in Illustris-TNG are found to be stronger in spheroids of TNG simulations while this is not significant in MassiveBlack-II, likely due to the much lower fraction of disk galaxies in this simulation.

The results from this study should help to inform the halo model approaches to model galaxy alignments on small scales. In future work, we plan to incorporate mock galaxy alignments in Euclid Flagship simulation \citep{2017ComAC...4....2P}, taking into account the mass and radial dependence of satellite alignments in hydrodynamic simulations. 

\section*{Acknowledgments}
AT acknowledges funding from Enabling Weak lensing Cosmology (EWC) through European Union’s Horizon 2020 research and innovation programme under grant agreement No 776247. TK is supported by a Royal  Society  University  Research Fellowship and EWC. AT thanks Rachel Mandelbaum and Henk Hoekstra for feedback on this work.
TDM acknowledges funding from NSF ACI-1614853, NSF AST-1517593, NSF AST-1616168 and NASA ATP 19-ATP19-0084.
TDM also acknowledges funding from  NASA ATP 80NSSC18K101, and NASA ATP NNX17AK56G.

\section*{Data availability}
The data underlying this article were accessed from the MassiveBlack-II \footnote{\url{https://www.mbii.phys.cmu.edu/data/}} and Illustris-TNG simulation database \footnote{\url{https://www.tng-project.org/data/}}. The derived data generated in this research will be shared on reasonable request to the corresponding author.

\bibliographystyle{mnras} \bibliography{draft}

\appendix

\section{Projection effects on Satellite shape - Central position correlation functions} \label{app:projection}

\begin{figure*} 
\begin{center}
  \includegraphics[width=3.2in]{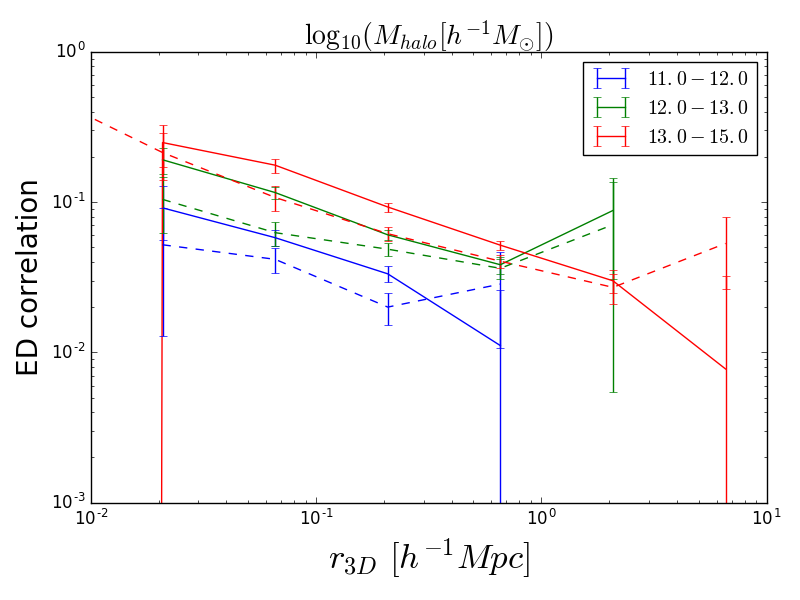}
  \includegraphics[width=3.2in]{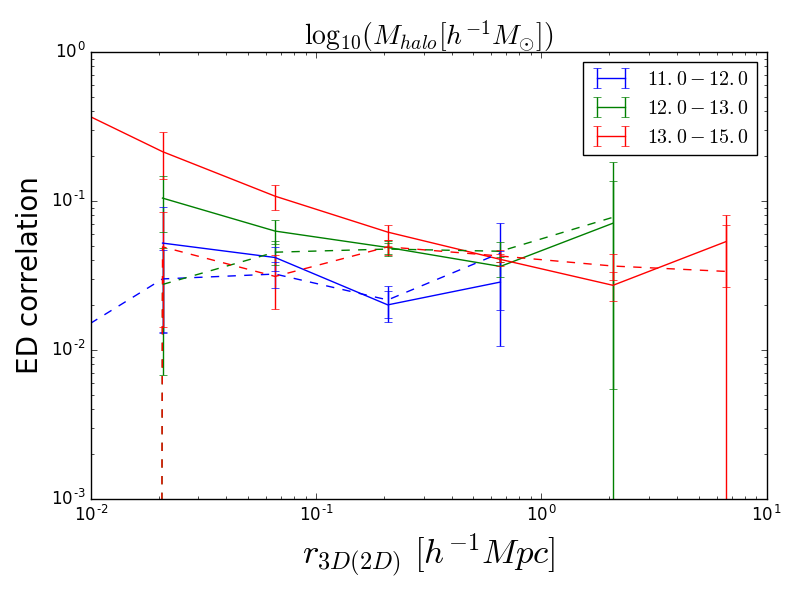}
\caption{\label{F:fig_ed_mb2_satshape_r3d2d} {\textit{Left:}} Ellipticity-direction (ED) correlation function of the $\rm{3D}$ and $\rm{2D}$ (projected) shape of the satellite galaxies with the location of the central galaxy within the given halo. The solid and dashed lines depict the ED correlation for the 3D shape and $\rm{2D}$ (projected) shapes respectively. {\textit{Right:}} ED correlation of $\rm{2D}$ (projected) shape of satellite galaxies with distances measured in $\rm{3D}$ and $\rm{2D}$. The solid and dashed lines depict the distances measured in 3D ($r_{3D}$) and 2D ($r_{2D}$) respectively. The correlation functions are plotted in halos of mass bins, $10^{11-12}h^{-1}M_{\odot}$, $10^{12-13}h^{-1}M_{\odot}$ and $10^{13-15}h^{-1}M_{\odot}$ at redshift, $z=0.06$.}
\end{center}
\end{figure*}

\begin{figure*} 
\begin{center}
\includegraphics[width=3.2in]{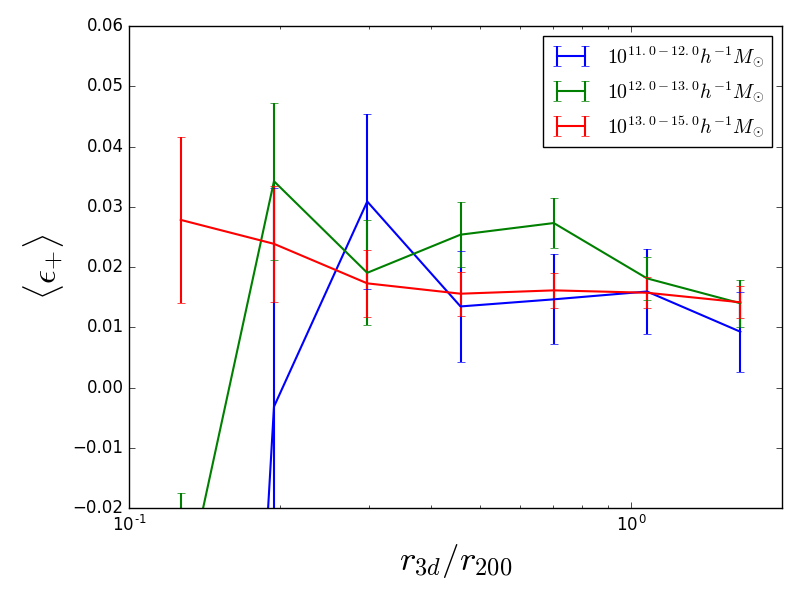}
\includegraphics[width=3.2in]{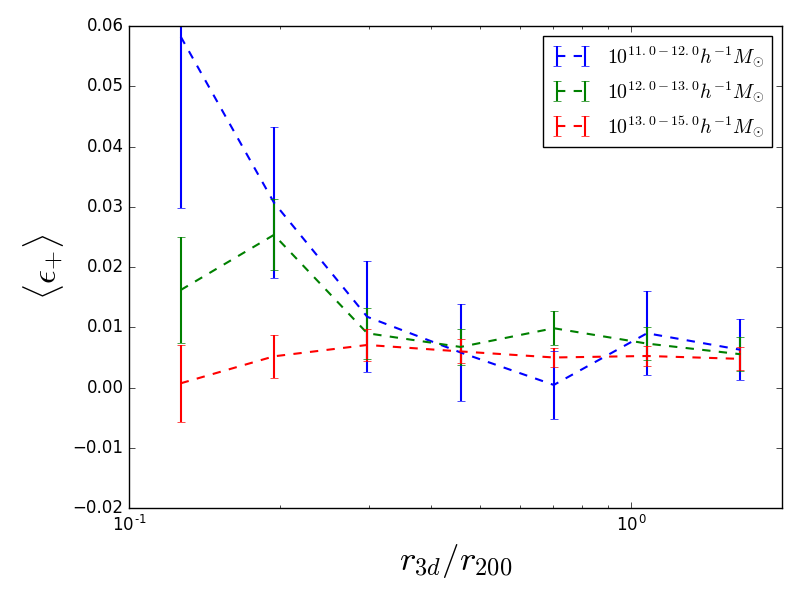}
\caption{\label{F:fig_avg_ellip_3d} The mean of the radial component of the projected ellipticities of the shapes of the stellar component of the satellite galaxies with respect to the host central galaxy. Here, the binning is in $\frac{r_{3d}}{r_{200}}$, the ratio of the $\rm{3D}$ distance of the satellite to the central galaxy, $r_{3d}$ and the $r_{200}$ of the central galaxy. The mean radial ellipticities are shown for the MassiveBlack-II ({\textit{Left}}) and Illustris-TNG ({\textit{Right}}) simulations in halo mass bins of $10^{11-12}h^{-1}M_{\odot}$, $10^{12-13}h^{-1}M_{\odot}$ and $10^{13-15}h^{-1}M_{\odot}$}
\end{center}
\end{figure*}

Here, we explore the effects of projected shapes and projected distances on the radial dependence of satellite alignments in more detail. In the left panel of Figure~\ref{F:fig_ed_mb2_satshape_r3d2d}, we plot the ED correlation of the $\rm{3D}$ stellar shapes (shown by solid lines) and projected ($\rm{2D}$) shapes (shown by dashed lines) of the satellite galaxies with the location of the central galaxies as a function of the $\rm{3D}$ distance. We can observe a slight decrease in the amplitude of alignment when using projected shapes. However, the radial scaling is similar to $\rm{3D}$ shapes. In the right panel of Figure~\ref{F:fig_ed_mb2_satshape_r3d2d}, we plot the alignment of the projected shapes, but the distances are binned in $\rm{3D}$ (shown by solid lines) and $\rm{2D}$ (shown by dashed lines). When binned in $\rm{2D}$ distances, we can observe a loss of radial dependence in the satellite alignment. In Figure~\ref{F:fig_avg_ellip_3d}, we plot the mean radial ellipticities of the projected shapes with distances binned in $r_{3{\rm d}}/r_{200}$ in MassiveBlack-II and IllustrisTNG simulations. In MassiveBlack-II , the decreasing radial dependence of the signal is seen in the highest mass bin, and weakly in the lower mass bins too. The TNG simulations show a radial dependence in the lower mass bins, but not in the highest mass bin which is also the case in $3\rm{D}$ satellite alignments as discussed before in Section~\ref{S:1hcorr}. So, we can conclude that the lack of radial dependence in the mean projected ellipticities of the satellite galaxies in simulations is mainly due to the effects of binning in projected distances  and to some extent due to normalization with $r_{200}$. The projected satellite shapes still tend to point towards the central galaxies with a stronger alignment at smaller distances. 

\section{ED correlation functions of centrals with satellites within $r_{200}$} \label{app:ed_inrad}

\begin{figure*} 
\begin{center}
  \includegraphics[width=3.2in]{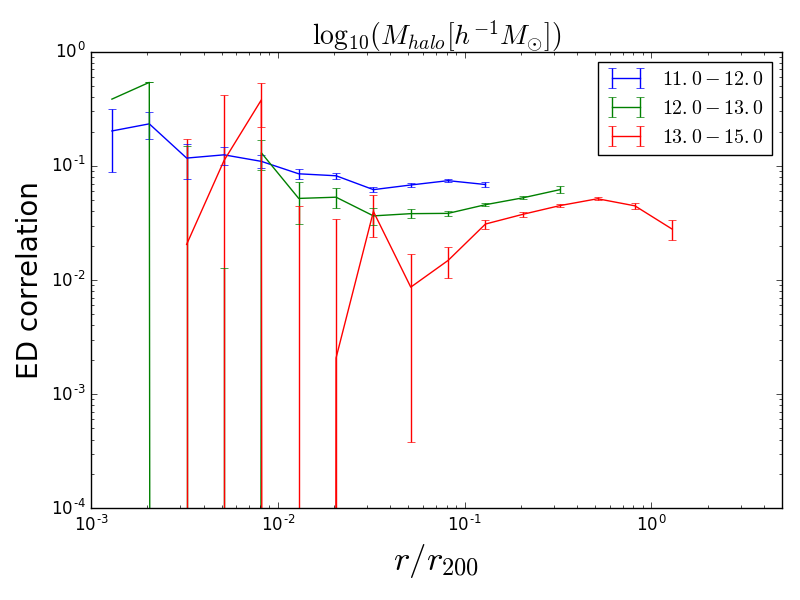}
  \includegraphics[width=3.2in]{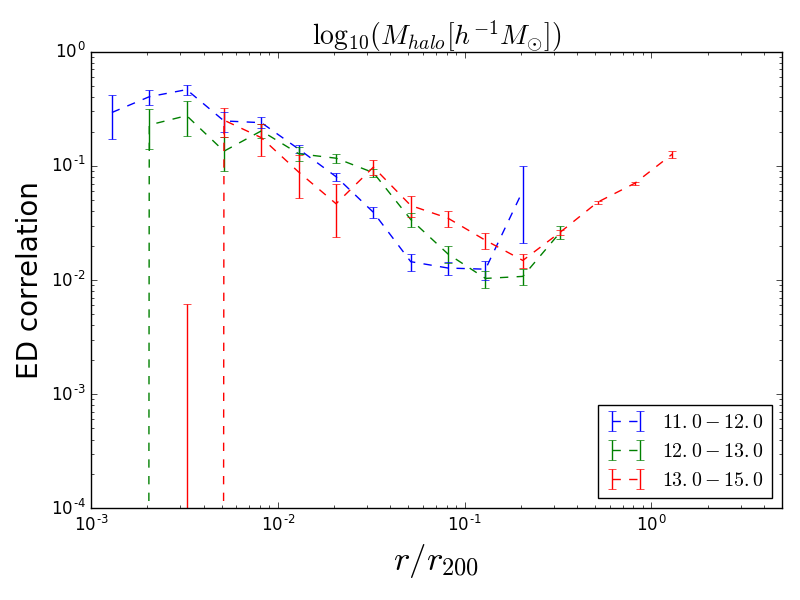}
\caption{\label{F:fig_ed_cen_sat_inradr200} Ellipticity-direction (ED) correlation function of the $3\rm{D}$ shape of the central galaxy with the location of the satellite galaxies within the distance, $r_{200}$ of the given halo in MassiveBlack-II ({\textit{Left}}) and Illustris-TNG ({\textit{Right}}) simulations. The correlation functions are plotted in halos of mass bins, $10^{11-12}h^{-1}M_{\odot}$, $10^{12-13}h^{-1}M_{\odot}$ and $10^{13-15}h^{-1}M_{\odot}$ at redshift, $z=0.06$.}
\end{center}
\end{figure*}

In Figure~\ref{F:fig_ed_cen_sat_inradr200}, we plot the ED correlation function of the central galaxy shape with position of satellite galaxies, including satellites only within the distance of $r_{200}$ for the given halo in the density sample. This is to understand if the increasing trend of the correlation function on large scales is seen when only satellite galaxies within $r_{200}$ are considered. For the highest mass bin, we can clearly see that the trend is similar to what we find using all satellites within the halo. This is consistent with our earlier discussion in Section~\ref{S:1hcorr} that the increase in correlation is due to the satellites on outer regions of halo tracing the shape of the dark matter halo. 

\section{Small-scale correlation functions in Illustris-TNG300 simulation} \label{app:tng300}
\begin{figure*}
\begin{center}
  \includegraphics[width=3.2in]{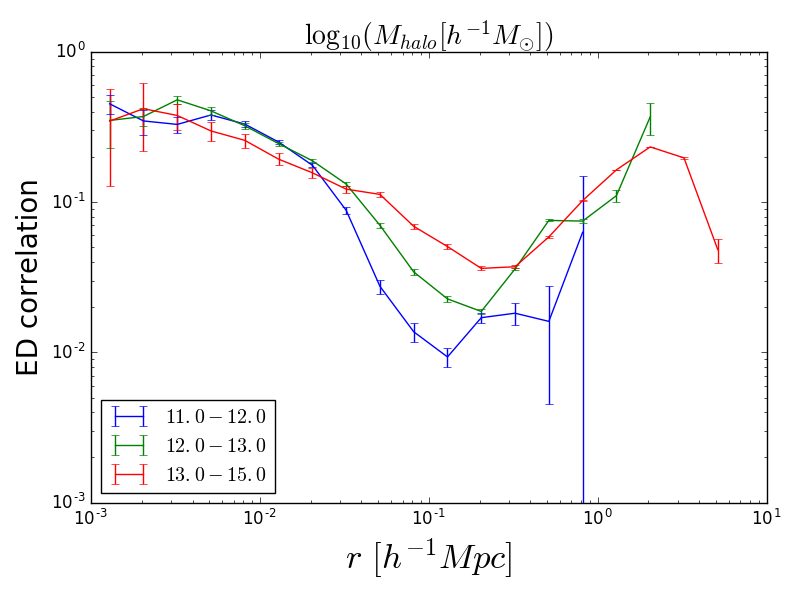}
  \includegraphics[width=3.2in]{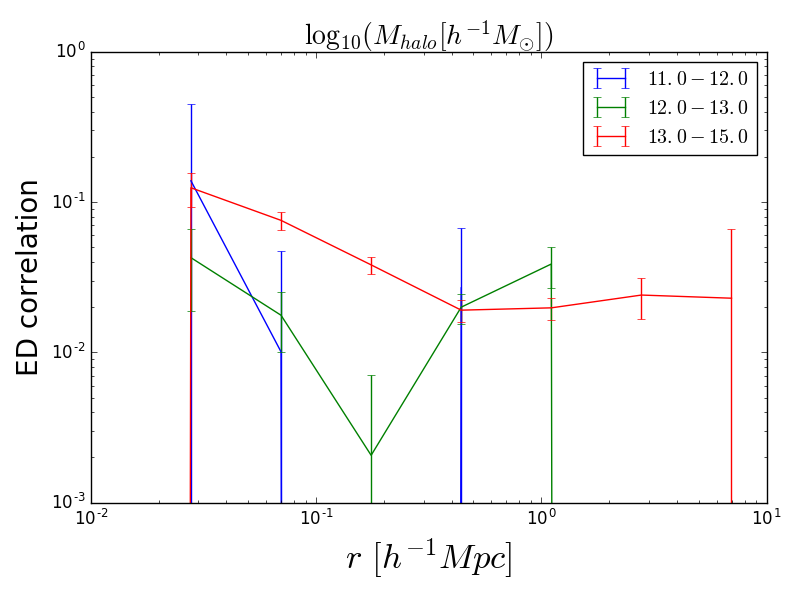}
\caption{\label{F:fig_ed_cen_sat_tng300} {\textit{Left:}} Ellipticity-direction (ED) correlation function of the $3\rm{D}$ shape of the central galaxy with the location of the satellite galaxies within the given halo in Illustris-TNG300 simulation({\textit{Left}}). {\textit{Right:}} Ellipticity-direction (ED) correlation function of the $3D$ shape of the satellite galaxies with the location of the central galaxy within the given halo. The correlation functions are plotted in halos of mass bins, $10^{11-12}h^{-1}M_{\odot}$, $10^{12-13}h^{-1}M_{\odot}$ and $10^{13-15}h^{-1}M_{\odot}$ at redshift, $z=0.06$.}
\end{center}
\end{figure*}

\begin{figure*} 
\begin{center}
\includegraphics[width=3.2in]{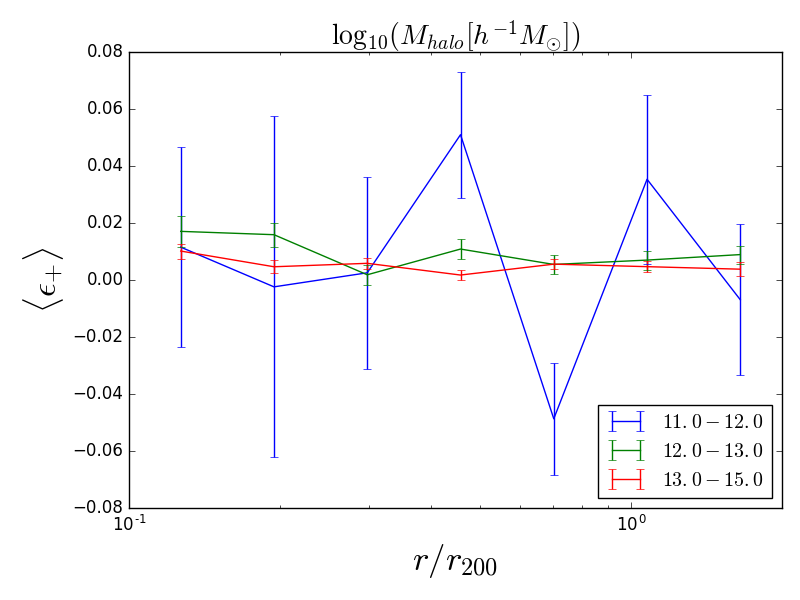}
\includegraphics[width=3.2in]{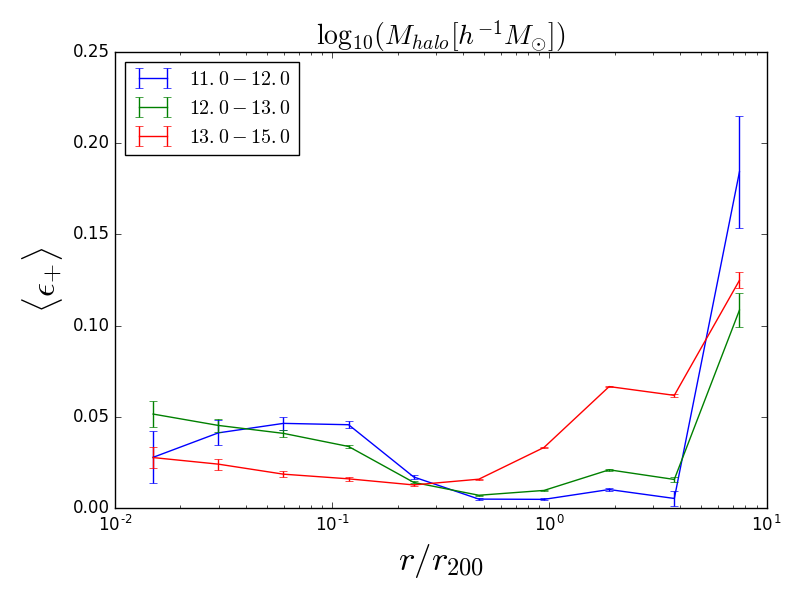}
\caption{\label{F:fig_avg_ellip_tng300} {\textit{Left:}} The mean of the radial component of the projected ellipticities of the shapes of the stellar component of the satellite galaxies with respect to the host central galaxy.The mean radial ellipticities are shown for the Illustris-TNG300 simulation in halo mass bins of $10^{11-12}h^{-1}M_{\odot}$, $10^{12-13}h^{-1}M_{\odot}$ and $10^{13-15}h^{-1}M_{\odot}$. {\textit{Right:}} Mean radial ellipticity of the projected shape of the central galaxy with the location of the satellite galaxies within the given halo.}
\end{center}
\end{figure*}

\begin{figure*} 
\begin{center}
\includegraphics[width=2.25in]{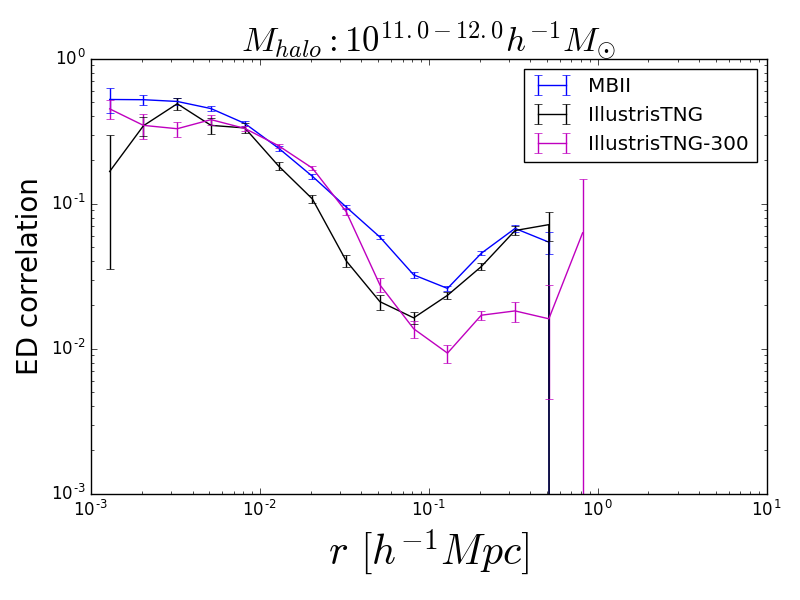}
\includegraphics[width=2.25in]{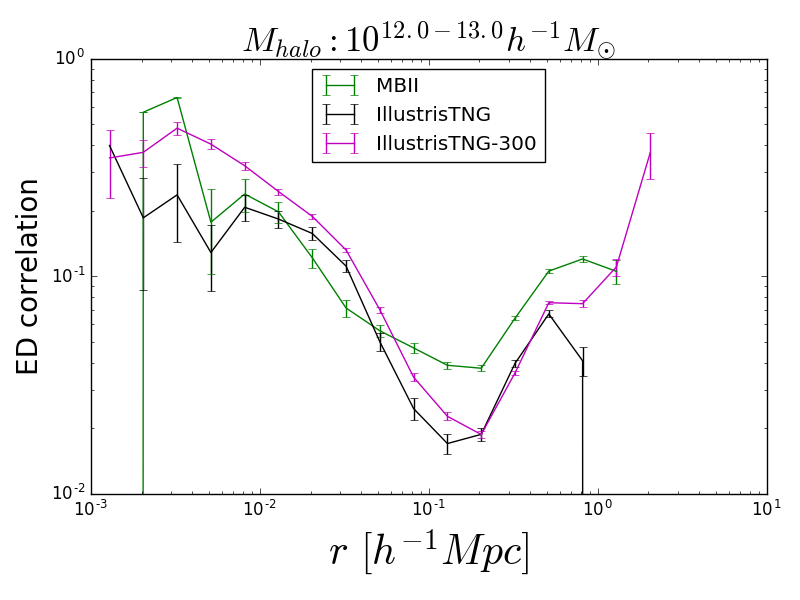}
\includegraphics[width=2.25in]{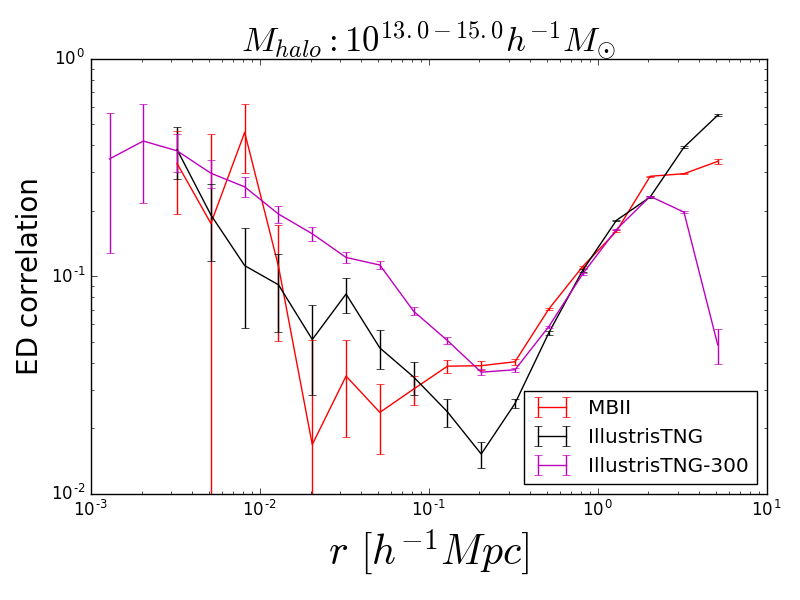}
\caption{\label{F:fig_ed_censhape_satpos_mb2tngs} {\textit{Left:}} The mean of the radial component of the projected ellipticities of the shapes of the stellar component of the satellite galaxies with respect to the host central galaxy.The mean radial ellipticities are shown for the Illustris-TNG300 simulation in halo mass bins of $10^{11-12}h^{-1}M_{\odot}$, $10^{12-13}h^{-1}M_{\odot}$ and $10^{13-15}h^{-1}M_{\odot}$. {\textit{Right:}} Mean radial ellipticity of the projected shape of the central galaxy with the location of the satellite galaxies within the given halo.}
\end{center}
\end{figure*}

\begin{figure*} 
\begin{center}
  \includegraphics[width=2.25in]{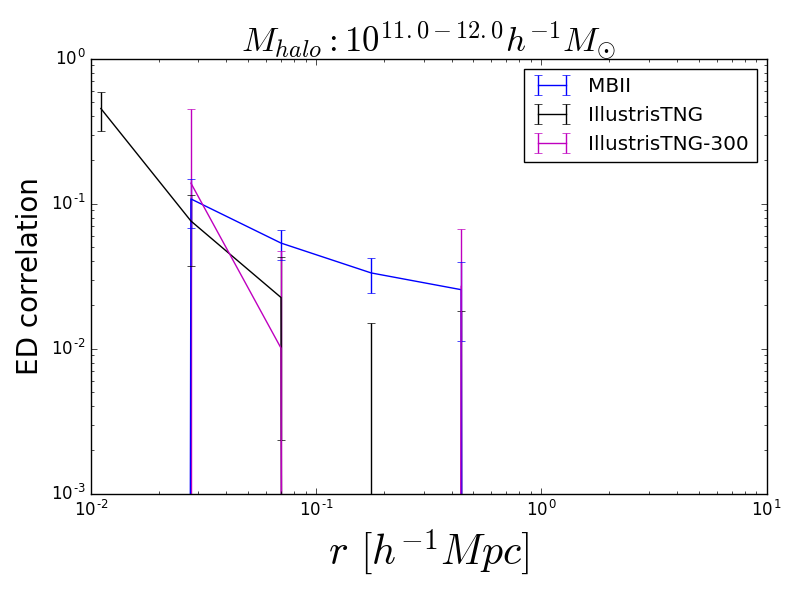}
  \includegraphics[width=2.25in]{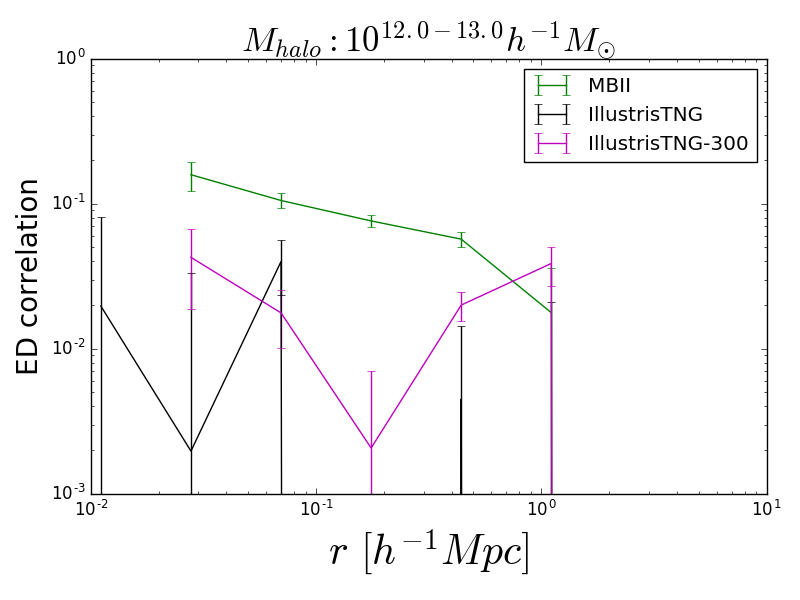}
  \includegraphics[width=2.25in]{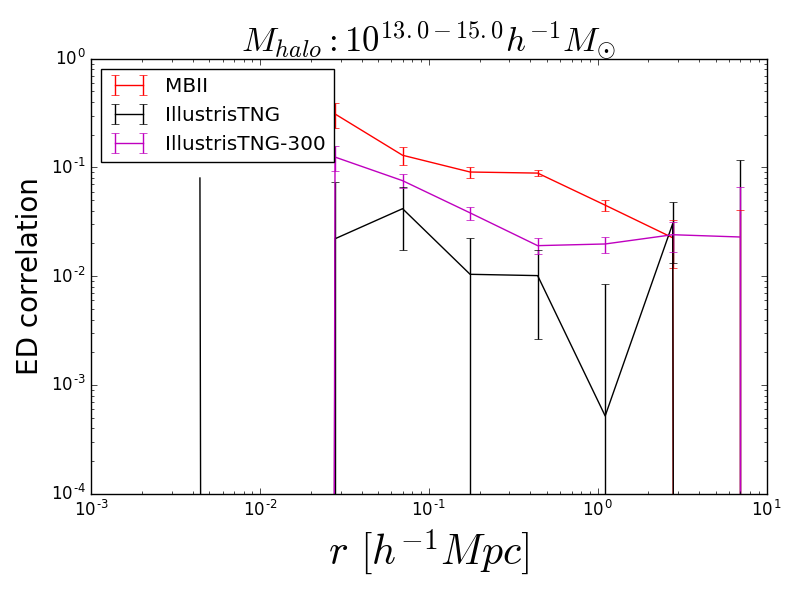}
\caption{\label{F:fig_ed_satshape_cenpos_mb2tngs} {\textit{Left:}} Ellipticity-direction (ED) correlation function of the $3\rm{D}$ shape of the central galaxy with the location of the satellite galaxies within the given halo in Illustris-TNG300 simulation({\textit{Left}}). {\textit{Right:}} Ellipticity-direction (ED) correlation function of the $3D$ shape of the satellite galaxies with the location of the central galaxy within the given halo. The correlation functions are plotted in halos of mass bins, $10^{11-12}h^{-1}M_{\odot}$, $10^{12-13}h^{-1}M_{\odot}$ and $10^{13-15}h^{-1}M_{\odot}$ at redshift, $z=0.06$.}
\end{center}
\end{figure*}

\begin{figure*} 
\begin{center}

  \includegraphics[width=2.25in]{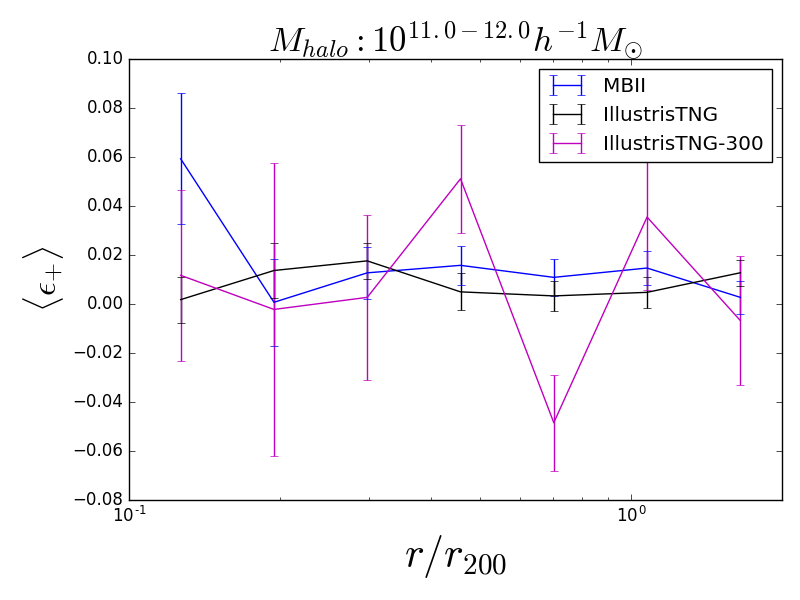}
  \includegraphics[width=2.25in]{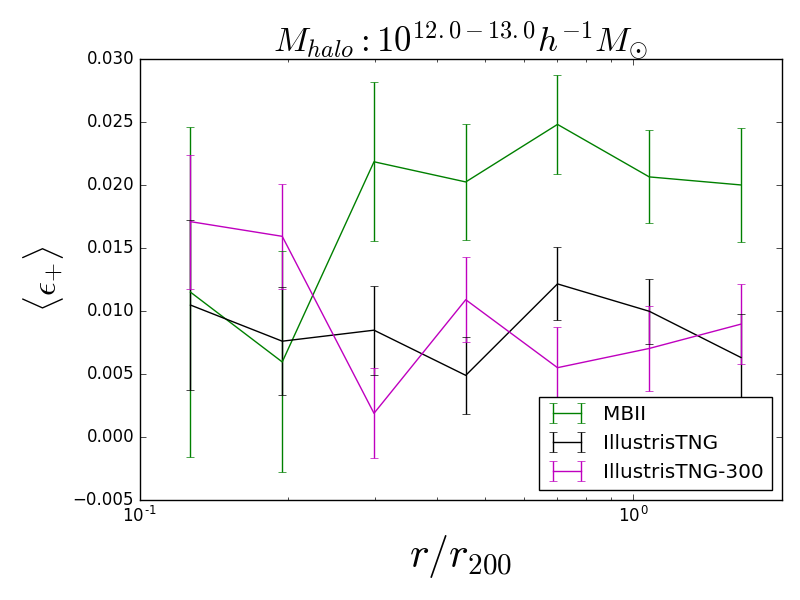}
  \includegraphics[width=2.25in]{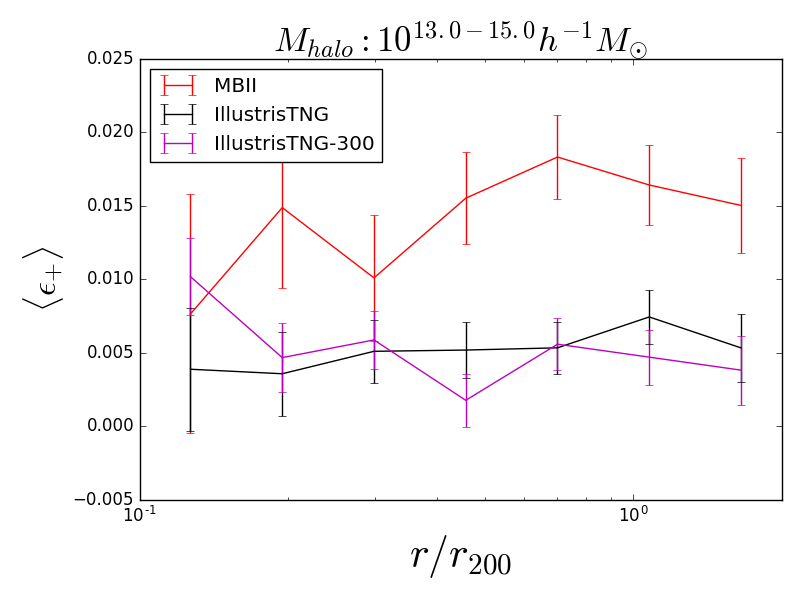}
\caption{\label{F:fig_avg_satellip_mb2tngs} {\textit{Left:}} The mean of the radial component of the projected ellipticities of the shapes of the stellar component of the satellite galaxies with respect to the host central galaxy. The mean radial ellipticities are shown for the Illustris-TNG300 simulation in halo mass bins of $10^{11-12}h^{-1}M_{\odot}$, $10^{12-13}h^{-1}M_{\odot}$ and $10^{13-15}h^{-1}M_{\odot}$. {\textit{Right:}} Mean radial ellipticity of the projected shape of the central galaxy with the location of the satellite galaxies within the given halo.}
\end{center}
\end{figure*}

\begin{figure*} 
\begin{center}
  \includegraphics[width=2.25in]{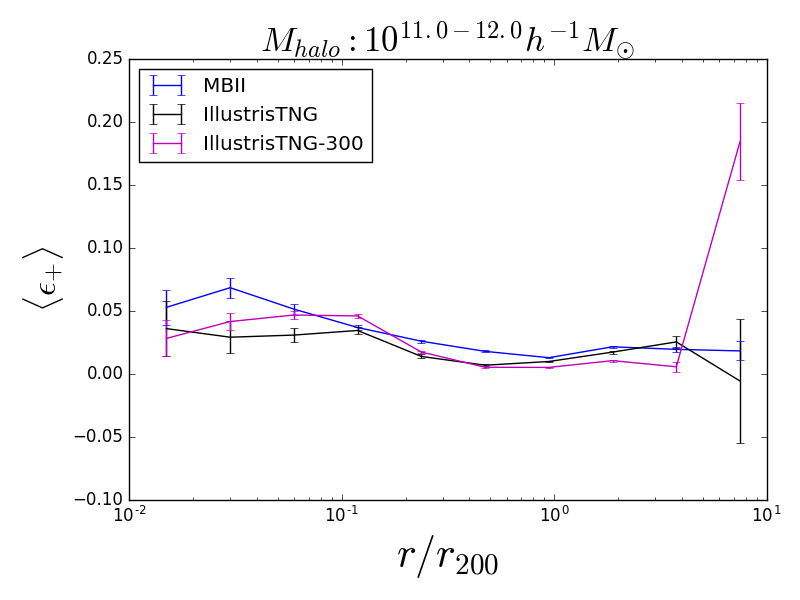}
  \includegraphics[width=2.25in]{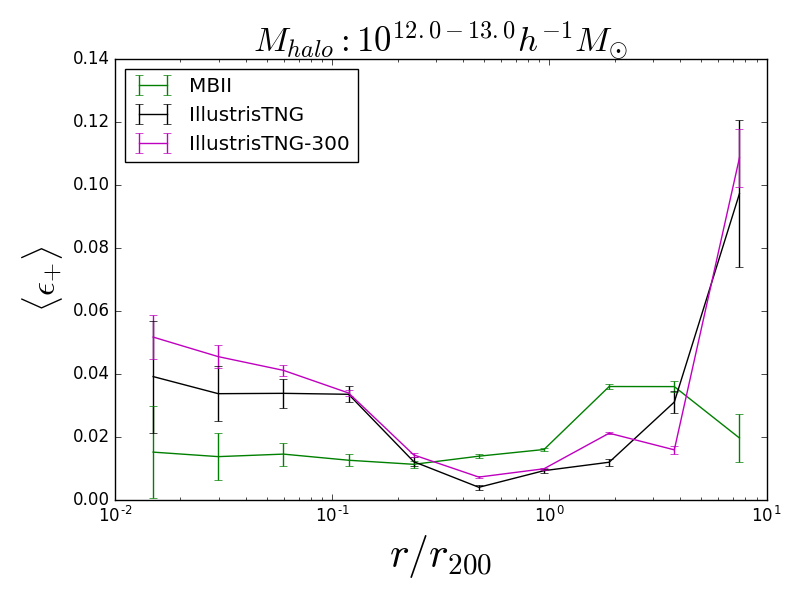}
  \includegraphics[width=2.25in]{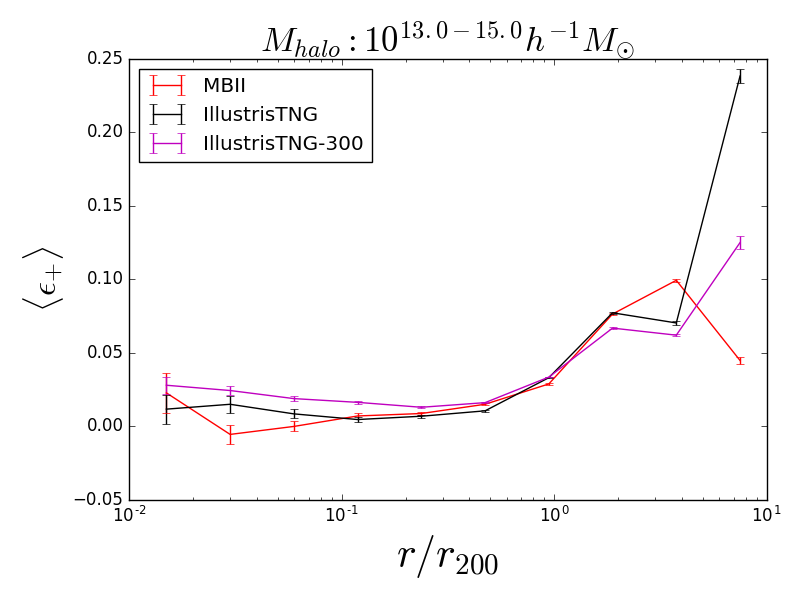}
\caption{\label{F:fig_avg_cenellip_mb2tngs} {\textit{Left:}} Ellipticity-direction (ED) correlation function of the $3\rm{D}$ shape of the central galaxy with the location of the satellite galaxies within the given halo in Illustris-TNG300 simulation({\textit{Left}}). {\textit{Right:}} Ellipticity-direction (ED) correlation function of the $3D$ shape of the satellite galaxies with the location of the central galaxy within the given halo. The correlation functions are plotted in halos of mass bins, $10^{11-12}h^{-1}M_{\odot}$, $10^{12-13}h^{-1}M_{\odot}$ and $10^{13-15}h^{-1}M_{\odot}$ at redshift, $z=0.06$.}
\end{center}
\end{figure*}

Here, we also extend the small-scale correlation functions to the larger volume Illustris-TNG300 simulation of boxsize $205 h^{-1}\rm{Mpc}$ at $z=0.06$. In Figure~\ref{F:fig_ed_cen_sat_tng300}, we plot the ED correlation function of the the alignment of the shape of the central galaxies with the location of satellites ({\textit{Left}} panel) and also the satellite alignment with central galaxy position ({\textit{Right}} panel) in the halo mass bins $10^{11-12}h^{-1}M_{\odot}$, $10^{12-13}h^{-1}M_{\odot}$ and $10^{13-15}h^{-1}M_{\odot}$. Figure~\ref{F:fig_avg_ellip_tng300} shows the projected mean ellipticities of the central shapes and satellite shapes with respect to the position of satellite and central galaxies respectively. In general, the mass dependent trend of all the correlations function is similar to those of Massive-Black II and TNG100 simulations discussed in Section~\ref{S:1hcorr}. However, for the alignments of the $\rm{3D}$ shapes of satellite galaxies with respect to the central ({\textit{Right}} panel of Figure~\ref{F:fig_ed_cen_sat_tng300}), we find a radially decreasing correlation function only in the highest mass bin. This is possibly due to fewer satellites in the lowest mass bin of TNG300 simulation, as the resolution is smaller when compared with TNG100. In Figures~\ref{F:fig_ed_censhape_satpos_mb2tngs},~\ref{F:fig_ed_satshape_cenpos_mb2tngs},~\ref{F:fig_avg_satellip_mb2tngs} and \ref{F:fig_avg_cenellip_mb2tngs}, we compare the alignment signals of the small-scale correlation functions in TNG300 simulation with those of TNG and Massive-Black II results within the same halo mass bins. Based on the ED correlation functions of the central galaxy shape with the satellite position shown in Figure~\ref{F:fig_ed_censhape_satpos_mb2tngs}, we can see that the amplitude of correlation function is higher for the central galaxies in TNG300 simulation, when compared to the galaxies in TNG100 in all mass bins while the radial scaling is similar. It is likely that the smaller resolution in the TNG300 simulation, box-size effects can lead to different galaxy properties and alignments although the baryonic physics model is the same. We have separately verified that the disk galaxy fraction is smaller in TNG300 simulation at all masses and the change is morphological fraction can be one of the reasons for stronger alignments in TNG300.  
The correlation function in Massive-Black II have larger amplitudes in the two lower mass bins, while the amplitude is similar in the highest mass bin. As discussed in Section~\ref{S:satsystem}, the satellite anisotropy is larger in MassiveBlack-II due to the stronger alignment of the shape of the stellar component with that of the dark matter component. Comparing the $\rm{3D}$ satellite alignments in Figure~\ref{F:fig_ed_satshape_cenpos_mb2tngs}, the amplitude of alignments are larger in the Massive-Black II simulation for all mass bins. In the TNG300 simulation, we can observe a significant satellite alignment signal only in the highest mass bin, where the alignment is higher than TNG100. The lack of resolution leads to no sufficient number of satellites in the lowest mass bins. The mean projected ellipticities of the satellite shapes with central positions are compared in Figure~\ref{F:fig_avg_satellip_mb2tngs}. Similar to satellite alignments in $\rm{3D}$, the amplitude of mean projected ellipticities are larger in Massive-Black II. Similar to Massive-Black II, we do not observe a radial dependence of the projected satellite alignment in both TNG100 and TNG300 simulations. As seen from Figure~\ref{F:fig_avg_cenellip_mb2tngs}, the projected ellipticities of the central galaxies are similar in the lowest and highest halo mass bins. The correlation functions for Massive-Black II galaxies in the intermediate mass bin shows a smaller amplitude on small scales, but higher on large scales when compared to those of TNG simulations.

\end{document}